\documentclass[aps,prd,amsmath,amssymb,preprint,superscriptaddress,longbibliography]{revtex4-1}

\usepackage[utf8]{inputenc}
\usepackage[T1]{fontenc}
\usepackage{graphicx}
\usepackage[dvipsnames]{xcolor}
\usepackage{dcolumn}
\usepackage{bm}
\usepackage[colorlinks=true, linkcolor=RoyalBlue, citecolor=RoyalBlue]{hyperref}
\usepackage[per-mode=symbol, separate-uncertainty]{siunitx}
\usepackage[capitalise]{cleveref}
\usepackage{enumerate}
\usepackage{float}
\usepackage{lineno}
\begin{document}

\title{Magneto-optics in a van der Waals magnet tuned by self-hybridized polaritons} 

\author{Florian Dirnberger$^\S$}
\email{fdirnberger@ccny.cuny.edu}
\affiliation{Department of Physics, City College of New York, New York, NY 10031, USA}
\thanks{Authors contributed equally.}

\author{Jiamin Quan$^\S$}
\affiliation{Department of Physics, The Graduate Center, City University of New York, New York, NY 10016, USA}
\affiliation{Photonics Initiative, CUNY Advanced Science Research Center, New York, NY, 10031, USA}
\affiliation{Department of Electrical Engineering, City College of the City University of New York, New York, NY, 10031, USA}
\thanks{Authors contributed equally.}

\author{Rezlind Bushati}
\affiliation{Department of Physics, City College of New York, New York, NY 10031, USA}
\affiliation{Department of Physics, The Graduate Center, City University of New York, New York, NY 10016, USA}

\author{Geoffrey Diederich}
\affiliation{Intelligence Community Postdoctoral Research Fellowship Program, University of Washington, Seattle, WA, USA}
\affiliation{Department of Physics, Department of Materials Science and Engineering, University of Washington, Seattle, WA, USA}

\author{Matthias Florian}
\affiliation{Department of Electrical and Computer Engineering, Department of Physics, University of Michigan, Ann Arbor, Michigan 48109, United States}

\author{Julian Klein}
\affiliation{Department of Materials Science and Engineering, Massachusetts Institute of Technology, Cambridge, Massachusetts 02139, USA}

\author{Kseniia Mosina}
\author{Zdenek Sofer}
\affiliation{Department of Inorganic Chemistry, University of Chemistry and Technology Prague, Technickaá 5, 166 28 Prague 6, Czech Republic}

\author{Xiaodong Xu}
\affiliation{Department of Physics, Department of Materials Science and Engineering, University of Washington, Seattle, WA, USA}

\author{Akashdeep Kamra}
\author{Francisco J. García-Vidal}
\affiliation{Departamento de Física Teórica de la Materia Condensada and Condensed Matter Physics Center (IFIMAC), Universidad Autónoma de Madrid, E- 28049 Madrid, Spain.}

\author{Andrea Al\`{u}}
\email{aalu@gc.cuny.edu}
\affiliation{Department of Physics, The Graduate Center, City University of New York, New York, NY 10016, USA}
\affiliation{Photonics Initiative, CUNY Advanced Science Research Center, New York, NY, 10031, USA}
\affiliation{Department of Electrical Engineering, City College of the City University of New York, New York, NY, 10031, USA}

\author{Vinod M. Menon}
\email{vmenon@ccny.cuny.edu}
\affiliation{Department of Physics, City College of New York, New York, NY 10031, USA}
\affiliation{Department of Physics, The Graduate Center, City University of New York, New York, NY 10016, USA}

%\date{\today}

\begin{abstract}
Controlling quantum materials with light is of fundamental and technological importance. 
By utilizing the strong coupling of light and matter in optical cavities~\cite{Garcia2021,Schlawin2022,Bloch2022}, recent studies were able to modify some of their most defining features~\cite{Sentef2018,Ashida2020,Appugliese2022}. 
In this work, we study the magneto-optical properties of a van der Waals magnet that supports strong coupling of photons and excitons even in the absence of external cavity mirrors.    
In this material -- the layered magnetic semiconductor CrSBr -- emergent light-matter hybrids called polaritons are shown to significantly increase the spectral bandwidth of correlations between the magnetic, electronic, and optical properties, enabling largely tunable optical responses to applied magnetic fields and magnons. 
Our results highlight the importance of exciton-photon self-hybridization in van der Waals magnets and motivate novel directions for the manipulation of quantum material properties by strong light-matter coupling.
\end{abstract}

\maketitle
%\begin{linenumbers}
%Introduction: 
Magnetic responses of optical excitations in solids are the key to efficiently interfacing magnetism and light, but materials supporting strong responses are rare.
It thus attracted considerable interest when studies recently demonstrated the exceptional magneto-optical properties of magnetic van der Waals (vdWs) crystals~\cite{Seyler2018,Zhang2019,Kang2020,Wilson2021,Klein2022-1}. 
In these layered materials, spin-related phenomena like the magneto-optical Kerr effect~\cite{Huang2017,Wu2019}, linear magnetic dichroism~\cite{Hwangbo2021}, and inherently polarized light emission~\cite{Seyler2018,Dirnberger2022} are often significantly enhanced in the spectral region of magnetic excitons -- an intriguing type of optical excitation formed by spin-polarized electronic states in magnets.
As a natural link between photons and spins, these magnetic excitons offer a unique opportunity to investigate the impact of strong exciton-photon coupling on the magneto-optical properties of layered magnetic systems. 

An archetypal vdWs material to leverage the effects of strong coupling is the antiferromagnetic semiconductor CrSBr.
Its optical spectrum supports pronounced excitonic signatures in the near-infrared region and moderately strong magnetic fields are sufficient to switch the equilibrium antiferromagnetic (AFM) order into a ferromagnetic (FM) configuration (see \cref{fig:1}a) below the Néel temperature $T_N=132\,K$~\cite{Wilson2021}, revealing an intimate relation between the electronic and the magnetic structure.
Opposed to magneto-optic effects that modulate the polarization of light, this relation directly alters the optical spectrum by modifying the exciton energy~\cite{Wilson2021}. 

Here we demonstrate the impact of strong light-matter coupling on the optical and magneto optical properties of the vdWs magnet CrSBr. 
Hybridization of magnetic excitons and photons is shown to fully determine the optical response of  mesoscopic crystals to applied magnetic fields and magnons.
We observe a significant enhancement of the spectral bandwidth of magneto-optic correlations due to the formation of tunable polaritons in crystals with and without external cavity mirrors. 
The excellent agreement between theoretical models and experimental observations highlights the virtually untapped potential of hybrid exciton-photon systems for magneto-optics.

\section*{Strong light-matter coupling in CrSBr}

\begin{figure*}[h!]
	\includegraphics[scale=0.9]{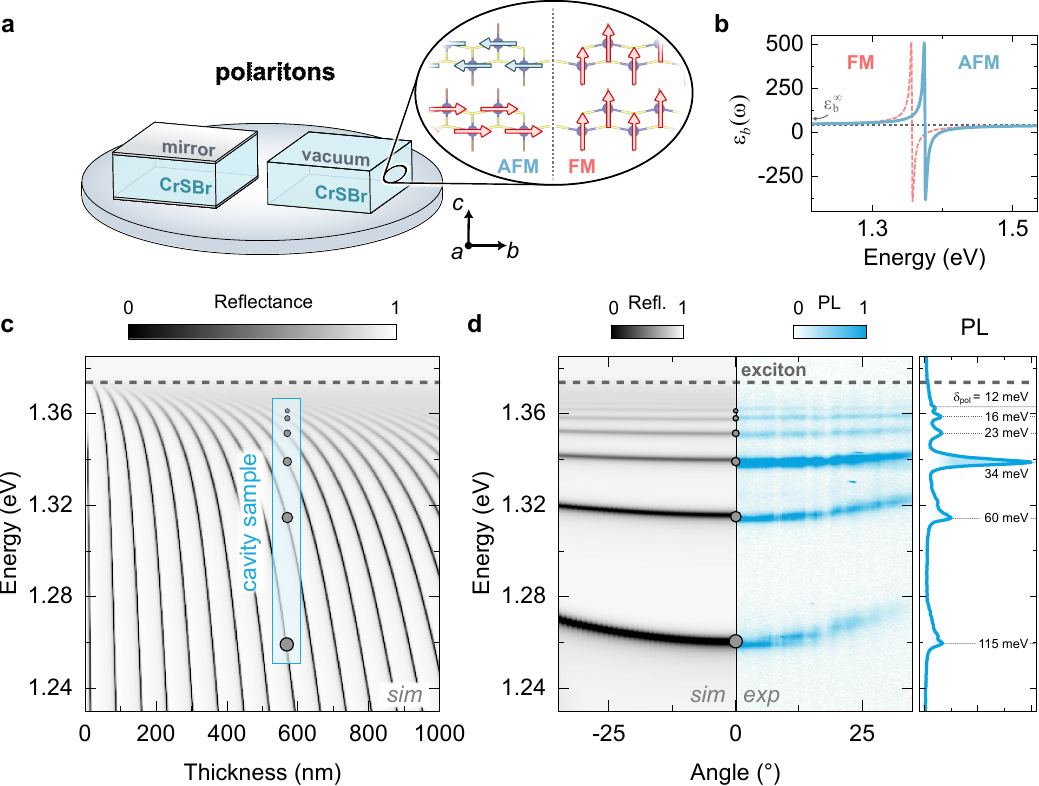}
	\caption{\textbf{Exciton-photon coupling in CrSBr cavities.} 
		\textbf{a}~Schematic illustrating two experimental approaches to support polaritons in mesoscopic CrSBr: bare crystals on a $\textrm{SiO}_2/\textrm{Si}$ substrate, and crystals embedded between highly reflective bottom and top mirrors. Magnified view: AFM and FM order ($\boldsymbol{B_{ext}}\parallel c$). Magnetic axes are the crystal $b$\,-- (easy), $a$\,-- (intermediate), and $c$\,-- (hard) axes (cf. Fig.~S1)~\cite{Wilson2021}. 
		\textbf{b}~Model dielectric functions of CrSBr derived in numerical calculations of optical reflectance spectra.  
		\textbf{c}~Simulation of the reflectance of CrSBr crystals for different thicknesses. Six prominent polariton branches are observed in a 580\,nm-thick sample with external cavity mirrors (gray circles). The exciton energy, $E_X=1.374$\,eV, determined from simulations, is indicated by the dashed gray lines. 
		\textbf{d}~Right panel: Angle-resolved and -integrated PL emission of a nominally 580\,nm-thick sample recorded at T=1.6\,K. The crystal was rotated by 45$^\circ$ with respect to the entrance slit of the spectrometer and the PL was analyzed with a polarizer along the $b$--axis. Left panel: Simulated reflectance map for conditions matching the experiment.
		\label{fig:1}}
\end{figure*}

Before discussing magneto-optic phenomena, we address the confinement of photons and their strong coupling with excitons in CrSBr crystals of mesoscopic size, i.e., for flake thicknesses varying from about 10 to 1000\,nm.
\Cref{fig:1}a depicts the two experimental configurations we use to confine optical modes.
One traps photons inside bare crystals due to the large dielectric mismatch at the interfaces to the environment, leading to pronounced self-hybridization effects~\cite{Canales2021}, while the other achieves larger photon confinement by adding highly reflective mirrors to the top and bottom surface. 
In both cases, we observe a strongly thickness-dependent series of optical states in the low-temperature optical reflectance, in stark contrast to the single exciton resonance at $\sim$1.34\,eV reported for bilayer samples~\cite{Wilson2021}. 
We are able to simulate the experimental reflectance signatures of different samples with and without external mirrors by modeling the dielectric function of CrSBr with just a single, strong oscillator to account for the main excitonic transition (see Figs.~\ref{fig:1}b\,\&\,S2-S4).
\Cref{fig:1}c shows the simulated reflectance as a function of crystal thickness for samples enclosed by external mirrors (cf. also polariton dispersion in bare crystals in Fig.~S2).

Our experiments and the theoretical models described in Section S2 unambiguously identify the optical states in mesoscopic CrSBr crystals as the hybrid exciton-photon quasiparticles known as exciton-polaritons.
Each state in the reflectance spectra corresponds to a specific branch of the polariton dispersion. 
Moreover, a comparison of our measurements with the model calculations demonstrates that the oscillator strength of excitons in CrSBr is amongst the largest known in solid-state systems, exceeding similar reports from gallium arsenide~\cite{Klingshirn2012}, transition-metal dichalcogenides~\cite{Munkhbat2018}, and hybrid perovskites~\cite{Fieramosca2018,Dang2020}. 
With Rabi splitting energies reaching $\hbar\Omega_R\approx0.24\,\textrm{eV}$, light-matter interactions in CrSBr are so pronounced, they are ascribed to the ultrastrong coupling regime (see Section S2\,\&\,Fig.~S5), which provides an intuitive explanation for the dramatically different optical spectra of mesoscopic crystals compared to bilayer samples.

Due to the pronounced self-hybridization effects already present in bare crystals, the addition of highly reflective cavity mirrors does not significantly change the polariton dispersion, but both our experiments and our simulations show they can further enhance experimental polariton signatures (see Figs.~S6\,\&\,7). 
\Cref{fig:1}d presents the angle-integrated and -resolved low-temperature photoluminescence (PL) emission of a 580\,nm-thick  flake with external cavity mirrors supporting six distinct polariton branches, all of which are in excellent agreement with the simulated and measured polariton reflectance spectra shown in Figs.~\ref{fig:1}d and S8.
A resonant state with small oscillator strength, potentially arising from defects~\cite{Klein2022-2}, enhances the intensity of the polariton peak near 1.34\,eV.
More importantly, we note an increasing contribution from photons in polariton branches with large zero-angle detuning $\delta_{pol}$ from the exciton resonance. 
Their stronger curvature directly evidences the reduction of the effective polariton mass due to the mixing with photons (see Fig.~S9). 
Overall, the different (average) exciton fractions \textbf{\textit{X}} of each polariton branch in this sample are well-suited to investigate the role of exciton-photon coupling in the magneto-optical responses of this material.

\section*{Tunable magneto-optic properties}

\begin{figure*}[h!]
	\includegraphics[scale=0.9]{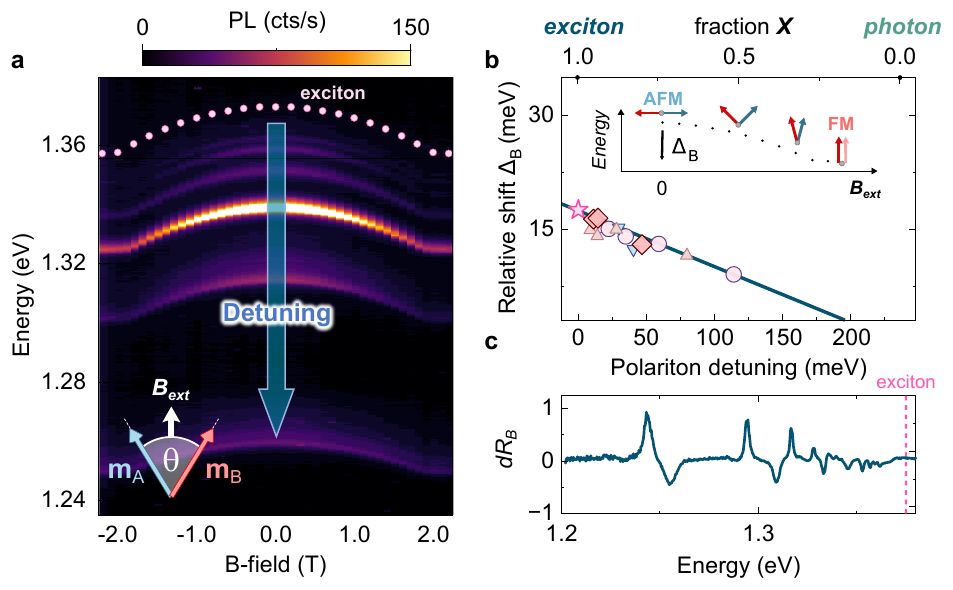}
	\caption{\textbf{Magnetic field response of polaritons.}
		\textbf{a}~Field-induced shift of the angle-integrated polariton PL emission. External field $\boldsymbol{B_{ext}}$ is applied along the $c$--direction. Pink dots show the field-dependence of excitons determined from numerical simulations and blue and red arrows represent sub-lattice magnetizations $m_A$ and $m_B$. Data was collected at T=1.6\,K from the sample shown in \cref{fig:1}d with a polarizer aligned along the $b$--axis.
		\textbf{b}~Maximum field-induced shift $\Delta_B$ as a function of detuning $\delta_{pol}$. Different symbols represent different samples with and without mirrors and same colors indicate polariton branches measured within each sample. Solid line shows the calculation $\Delta_B (\boldsymbol{X}) = \Delta_B \boldsymbol{X}$, where $\Delta_B = -17.5$\,meV (star symbol) is extracted from our numerical simulations. 
		Inset: Schematic of field-induced spin canting and the resulting shift $\Delta_B$.  
		\textbf{c}~Field-induced changes in the reflectance spectrum $dR_B$ of the sample shown in \textbf{a} ($B_{ext}=B_{sat}$). 
		\label{fig:2}}
\end{figure*}

Analogous to previous studies of the electric field response of polaritons~\cite{Lim2017,Bajoni2008}, we investigate the effects of an external magnetic field $\boldsymbol{B_{ext}}$. 
As described in Section S3, a field along the magnetic hard axis causes the canting of spins, which alters the angle $\theta$ between the two sub-lattice magnetizations and results in a decrease of exciton energies, $\Delta E = \Delta_B\cos^2(\theta/2)$~\cite{Wilson2021}. 
The PL emission of each branch in our 580\,nm sample therefore shifts towards lower energies, along a bell-shaped curve, until saturation occurs at $\boldsymbol{B_{ext}}=B_{sat}$ (see \cref{fig:2}a). 
We determine the maximum shift $\Delta_B (\boldsymbol{X})$ of polariton branches with different detuning $\delta_{pol}$ and exciton fraction $\boldsymbol{X}$ in several strongly coupled samples both with and without external mirrors and find that the magnetic field response of polaritons is strongly affected by hybridization.
\Cref{fig:2}b indicates a linear dependence of the magnetic shift on the exciton fraction $\boldsymbol{X}$, which modifies the dispersion and the effective mass of polariton branches (see Fig.~S9).
While the largest energy shifts are obtained for exciton-like branches, the strongest reflectance responses occur for more photon-like polaritons due to a decrease in absorption at lower energies (see \cref{fig:2}c). 
By transferring the correlation of the magnetic and the electronic structure to polariton states deep inside the optical gap, strong coupling substantially increases the bandwidth of the magneto-optical response of CrSBr to an applied field. 
Hence, even at energies much below the exciton resonance, where bi- and few layer samples are transparent, magnetic fields profoundly change the optical reflectance of mesoscopic CrSBr crystals by up to 90$\%$. 

\begin{figure*}[h!]
	\includegraphics[scale=0.9]{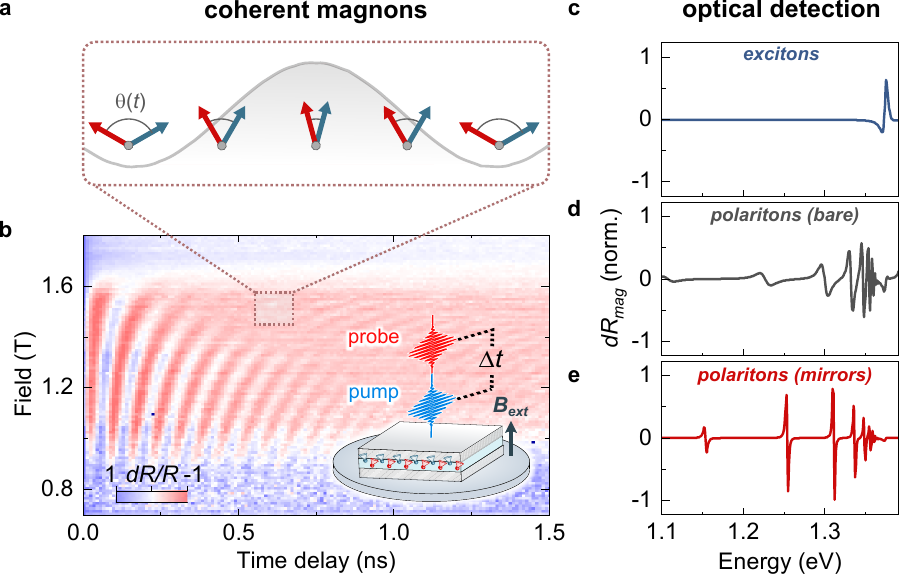}
	\caption{\textbf{Polariton-magnon coupling and coherent magnon effects.}
		\textbf{a}~Coherent magnon oscillations periodically modulate the angle $\theta(t)$ between the two sub-lattice magnetizations. 
		\textbf{b}~Energy-integrated pump-probe measurement of the transient reflectance of two closely spaced, exciton-like polariton branches ($\boldsymbol{X}\approx$\,0.9). Curve shows the normalized differential reflectance $dR/R$ recorded at different fields $\boldsymbol{B_{ext}}$ applied along the $c$--axis at $T$=\,50\,K. 
		\textbf{c-e}~Calculations of the magnon-induced reflectance changes $dR_{mag}$ based on the model described in Section S4B for excitons in a bilayer, and polaritons in 580\,nm thick crystals with and without external mirrors. Data normalized to the maximum in \textbf{e}. 
		\label{fig:3}}
\end{figure*}

Prominent magneto-optical effects can also be obtained by changing the internal magnetic order of CrSBr via magnons. 
Recent pump-probe studies demonstrate time-dependent oscillations of the exciton energy that arise from the periodic modulation of $\theta(t)$ by coherent magnons (cf. illustration in \cref{fig:3}a)~\cite{Bae2022,Geoff2022}. 
We observe similar effects for polaritons when we pump a thin crystal enclosed by external mirrors using ultrashort optical pulses, and integrate the transient reflectance of resonantly tuned probe pulses in energy. 
Subsequent to the ultrashort excitation, the polariton reflectance is strongly modulated, as shown in \cref{fig:3}b, by oscillations that excellently match the frequency-field dependence of coherent magnons in CrSBr (cf. Refs.~\cite{Bae2022,Geoff2022,Cham2022} and Fig.~S13). 
To highlight the differences between the magneto-optic response of excitons and polaritons, \cref{fig:3}d-f compares the calculated changes in the reflectance spectrum induced by coherent magnons.
Like in the response to a static field, polaritons substantially increase the spectral bandwidth of this magneto-optic effect in CrSBr.
Moreover, external cavity mirrors may significantly enhance the sensitivity of optical magnon detection, since the small polariton linewidth causes large relative changes in reflectance. 

\begin{figure*}[h!]
	\includegraphics[scale=0.9]{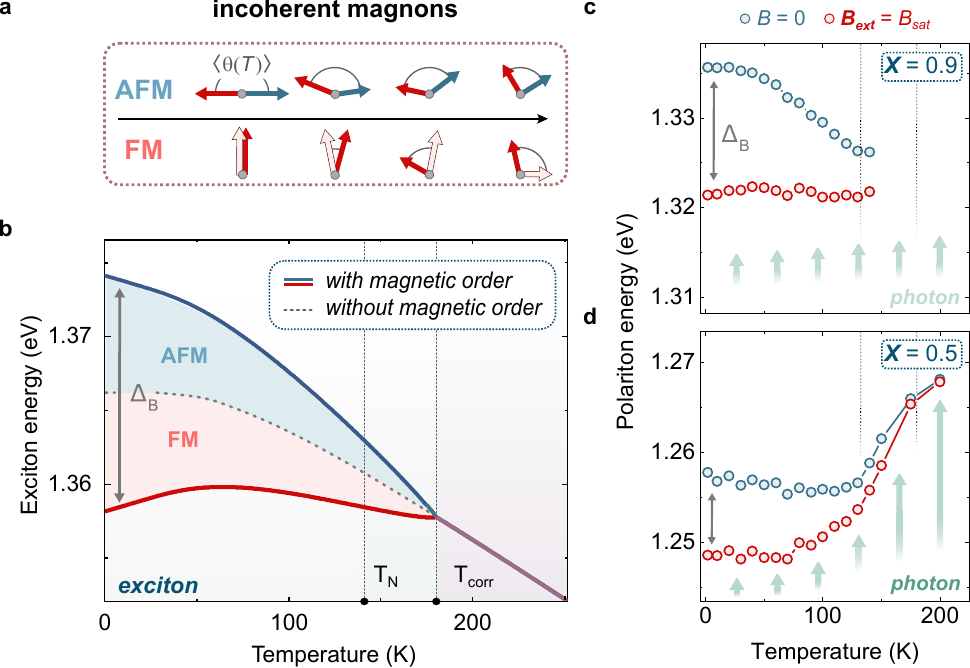}
	\caption{\textbf{Magneto-optical response of CrSBr to incoherent magnons.}
		\textbf{a}~At finite temperatures \textit{T}, incoherent magnons cause spins to fluctuate around their equilibrium position, changing the average angle $\langle\theta(\mathit{T})\rangle$ between sub-lattice magnetizations in AFM ($\langle\theta(\mathit{T})\rangle < \theta(0)$) and FM ($\langle\theta(\mathit{T})\rangle > \theta(0)$) configurations. FM order illustrated for $\boldsymbol{B_{ext}}\parallel c$.
		\textbf{b}~Model of the temperature dependence of excitons in CrSBr for initial AFM and FM order. Effects of the magnetic order strongly impact exciton energies up to a temperature $T_{corr}>T_N$.
		\textbf{c,d}~Temperature-dependent emission energy of polariton branches ($\boldsymbol{X}=0.9$ and $\boldsymbol{X}=0.5$) measured for initial AFM ($\boldsymbol{B_{ext}}=0$) and FM ($\boldsymbol{B_{ext}}\parallel c=B_{sat}$) configuration. The polarization was analyzed along the $b$--axis. Due to thermal broadening effects, exciton-like polaritons fade more rapidly from the spectrum. Large photon contributions can overcome the magnon- and phonon-induced shifts of excitons towards lower energies, facilitating a reversal of the temperature-dependent shift
		\label{fig:4}}
\end{figure*}

Beyond this magneto-optical response of CrSBr to coherent magnons, our study unveils a previously unobserved magneto-optic effect that does not require coherence. 
Magnons excited by non-zero temperatures completely lack the coherence imprinted by ultrafast optical excitation, but still change the average angle $\langle\theta(\mathit{T})\rangle$ (cf. \cref{fig:4}a), which in turn affects the exciton component of our polaritons due to the coupling of the electronic and the magnetic structure. 
Analytic magnon theory presented in Section S3 suggests that below $T_N$ the temperature dependence of excitons in CrSBr is primarily determined by the population of incoherent magnons.
Depending on whether the magnetic structure is initialized with AFM or FM order, incoherent magnons are expected to cause a shift that is linear at low temperatures ($T\ll T_N$) and either decreases or increases the energy of excitons. 
Because short-range correlations survive longer than the global magnetic ordering~\cite{Lopez2022,Liu2022}, the general effect extends even beyond the Néel temperature.
In CrSBr, the coupling of excitons to magnons thus adds to the typical phonon-related temperature dependence of the exciton energy known from conventional semiconductors~\cite{Odonnel1991}.
\Cref{fig:4}b shows the result of our exciton model that includes the effects of the magnetic order and the interaction of excitons with phonons, as described in Section S4. 

The qualitative agreement between this model and the measured temperature-field dependence of an exciton-like polariton branch with $\boldsymbol{X}=0.9$ shown in \cref{fig:4}c directly confirms our prediction of a magneto-optical response to incoherent magnons. 
Hence, transient exciton spectroscopy emerges as a novel optical probe to monitor the incoherent magnon population and temperature.

The hybridization with photons has a major effect on the thermal response of excitons. 
For $\boldsymbol{X}=0.5$, in \cref{fig:4}d, we observe a reversal of the polariton temperature dependence, resulting in a shift towards higher energies.
As the field-induced shift $\Delta_B$ vanishes in more photon-like polariton branches, effects of magnons and phonons on excitons are increasingly compensated by temperature-induced changes in the static refractive index that determine the energy of the photon component in mesoscopic CrSBr crystals.
Overall, our ability to predict this unique magneto-optical response (Section S4 \& Figs.~S14-S16), the impact of external fields, and the effects of magnons by numerical models, highlights the deterministic nature of engineering magneto-optical properties via strong exciton-photon coupling. 

\section*{Summary}
In conclusion, it is shown that strong exciton-photon coupling can determine the magneto-optical properties of mesoscopic van der Waals magnetic crystals. 
Hybrid light-matter excitations in the magnetic semiconductor CrSBr substantially increase the spectral bandwidth of magneto-optic responses by extending correlations between magnetic, electronic, and optical properties to states that lie deep within the band gap of the material. 
The major role of hybridization in the coupling of magnetism and light evidenced in our study encourages further research to pursue the modification of quantum materials by strong light-matter coupling.\\

%%%%%%%%%%%%%%%%%%%%%%%%%%%%%%%%%%%%%%%%

%References: 

%\bibliography{MS_CrSBr-mag-pol}
%\bibliographystyle{naturemag}

\section*{Methods}
\subsection{Crystal growth}
CrSBr bulk single crystals were synthesized by a chemical vapor transport method. 
Chromium (99.99\%, -60 mesh, Chemsavers Inc., USA), sulfur (99.9999\%, 1-6mm granules, Wuhan XinRong New Materials Co., China), and bromine (99.9999\%, Merck, Czech Republic) with a general stoichiometry of about 1:1:1 were placed in a quartz glass ampoule.
In fact, Sulfur and bromine were used with 2 atomic percentage excess over the stoichiometry. 
The ampule (250\,x\,50\,mm) was purged with argon and sealed under high vacuum while the charge was cooled with liquid nitrogen to avoid losses of bromine. 
The charge was first pre-reacted at 500~$^\circ$C, 600~$^\circ$C and 700~$^\circ$C for 10 hours at each temperature, while the top of the ampule was kept at 200~$^\circ$C. 
Finally the crystal growth was performed in a two-zone gradient furnace, where the charge temperature was gradually increased from 900~$^\circ$C to 920~$^\circ$C over a period of 6 days and the temperature of the growth zone was decreased from 900~$^\circ$C to 800~$^\circ$C. These temperatures of 920~$^\circ$C and 800~$^\circ$C were maintained for another 6 days.
Then the ampule was cooled to room temperature and opened under argon atmosphere in a glovebox. 

\subsection{Microcavity fabrication}
To fabricate bottom mirrors for our microcavities, highly reflective Bragg mirrors were grown by the plasma-enhanced chemical vapor deposition (PECVD) of 8 pairs of Silicon nitride/Silicon dioxide layers on Silicon wafers. 
First, millimeter-sized CrSBr crystals were thinned-down in multiple cycles of stick-and-release on blue tape (PVC tape 224PR, Nitto) before being transfered onto a polydimethylsiloxane (PDMS, AD series, Gel-Pak) film. 
By moderately pressing the PDMS film onto a Bragg mirror (or a $\textrm{SiO}_2/\textrm{Si}$ substrate, oxide thickness \SI{285}{\nm}) and swiftly releasing it, a multitude of crystals with typical thicknesses of tens to hundreds of nanometers were transfered. 
Then, a 35\,nm-thin layer of gold (or a highly reflective Bragg mirror) was deposited by electron-beam physical vapor deposition (or PECVD) to form a top mirror. 

\subsection{Optical spectroscopy}
For time-integrated spectroscopy measurements, samples were mounted inside a closed-cycle cryostat with a base temperature of 1.6\,K equipped with a 9\,T solenoid magnet.   
Samples can be mounted vertically, or horizontally to apply magnetic fields along in-plane or out-of-plane directions.
In all configurations, light is focused onto the sample at normal incidence. 

Reflectance spectra were obtained by focusing the attenuated output of a spectrally broadband tungsten-halogen lamp to a spot size of $\SI{2.0}{\um}$ by a $\SI{100}{\times}$ microscope objective (NA=0.81) mounted inside the cryostat. 
For polarization-resolved measurements, the reflected signal was analyzed by a combination of a half-waveplate and a linear polarizer along the axis of maximum PL intensity. 
Angle-integrated and angle-resolved measurements were recorded by respectively focusing either real-space images or the back-focal plane of the objective onto a spectrometer connected to a charge-coupled device (CCD). 
For the PL experiments, unless specified otherwise, a continuous-wave laser with 2.33\,eV energy and variable average output power was focused onto the sample to a spot size of $\SI{1.0}{\um}$ by the same objective used for reflectance spectroscopy. 
The collected PL signal was directed towards the spectrometer and spectrally filtered to remove the laser emission and, if required, analyzed by the polarization optics. 

Transient optical reflectivity measurements were performed by tuning the output of a titanium sapphire oscillator into resonance with the closely-spaced, highly exciton-like branches of a $\sim$100\,nm-thick CrSBr crystal embedded between two DBR mirrors. 
The output was frequency doubled and the second harmonic and fundamental were separated into pump and probe arms of the experiment by a dichroic mirror.
The probe beam was sent to a retroreflector mounted on a motorized translation stage in order to produce the pump-probe delay. 
Each beam was sent through a waveplate and polarizer to simultaneously attenuate the beams and align their polarization to the crystal axes. 
The beams were recombined and sent through a 0.6 NA microscope objective onto the sample. 
The back-reflected beam was measured on a photodiode with a lock-in amplifier demodulating the signal at the frequency of a mechanical chopper placed in the pump arm of the experiment. 
To produce the time-domain data, the delay stage was continuously swept at low speed while streaming data from the lock-in amplifier to the host computer at a high sampling rate (> 100 KHz), which produced time traces with 1 picosecond resolution in the data presented here. 
Multiple traces (4 < N < 25) were recorded and averaged, depending on the desired signal-to-noise ratio. 
The samples were kept at 50\,K, far below the Néel temperature of $\sim132$\,K, in an optical cryostat with an integrated vector magnet capable of applying fields along any arbitrary direction on the unit sphere.\\

\textbf{Acknowledgments:}

Work at CCNY was supported through the NSF DMR- 2130544 and the NSF CREST IDEALS center (V.M.M.), DARPA Nascent Light-Matter Interaction program (R.B.) and the German Research Foundation through project 451072703 (F.D.). 
A.A. and J.Q. have been supported by the Office of Naval Research, the Air Force Office of Scientific Research and the Simons Foundation. 
V.M.M and A.A acknowledge funding from the US Air Force Office of Scientific Research – MURI grant FA9550-22-1-0317. 
We acknowledge the use of the CUNY Advanced Science Research Center Nanofabrication Facility for the device fabrication.
Z.S. was supported by ERC-CZ program (project LL2101) from the Ministry of Education Youth and Sports (MEYS). 
The transient magneto-optical measurement at U. Washington is mainly supported by the Department of Energy, Basic Energy Sciences, Materials Sciences and Engineering Division (DE-SC0012509). 
This research was supported by an appointment to the Intelligence Community Postdoctoral Research Fellowship Program at University of Washington, administered by Oak Ridge Institute for Science and Education through an interagency agreement between the U.S. Department of Energy and the Office of the Director of National Intelligence.
M. F. acknowledges support by the Alexander von Humboldt foundation.
A.K. and F.J.G.-V. acknowledge support by the Spanish Ministry for Science and Innovation-Agencia Estatal de Investigación (AEI) through Grants PID2021-125894NB-I00 and CEX2018-000805-M (through the María de Maeztu program for Units of Excellence in R\&D) and by the Autonomous Community of Madrid, the Spanish government and the European Union through grant MRR Advanced Materials (MAD2D-CM).\\

\textbf{Author contributions:}

F.D. conceived the experimental idea and interpreted the results together with V.M.M. and all authors.
F.D., R.B., J.Q., and G.D. performed the experiments and conducted the data analysis, Z.S. and K.M. synthesized the CrSBr crystals, and A.K., F.J.G.V., and M.F. developed the analytic and numeric theories.  
F.D. wrote the manuscript with input from all authors and supervised the project with A.A. and V.M.M.\\

\textbf{Data availability:}
The main data sets generated and analyzed in this study are available
at [insert link in proof stage]. Supplementary data will be provided by the corresponding authors upon request.\\

\textbf{Competing interests:} 
The authors declare that they have no competing interests.\\

\textbf{Additional information:} 
Supplementary Information is available for this paper.
Correspondence and requests for materials should be addressed to Florian Dirnberger (fdirnberger@ccny.cuny.edu).\\
Reprints and permissions information is available at www.nature.com/reprints.\\

\end{document}

\title{SUPPLEMENTARY INFORMATION:\\Magneto-optics in a van der Waals magnet tuned by self-hybridized polaritons} 

\author{Florian Dirnberger$^\S$}
\email{fdirnberger@ccny.cuny.edu}
\affiliation{Department of Physics, City College of New York, New York, NY 10031, USA}
\thanks{Authors contributed equally.}

\author{Jiamin Quan$^\S$}
\affiliation{Department of Physics, The Graduate Center, City University of New York, New York, NY 10016, USA}
\affiliation{Photonics Initiative, CUNY Advanced Science Research Center, New York, NY, 10031, USA}
\affiliation{Department of Electrical Engineering, City College of the City University of New York, New York, NY, 10031, USA}
\thanks{Authors contributed equally.}

\author{Rezlind Bushati}
\affiliation{Department of Physics, City College of New York, New York, NY 10031, USA}
\affiliation{Department of Physics, The Graduate Center, City University of New York, New York, NY 10016, USA}

\author{Geoffrey Diederich}
\affiliation{Intelligence Community Postdoctoral Research Fellowship Program, University of Washington, Seattle, WA, USA}
\affiliation{Department of Physics, Department of Materials Science and Engineering, University of Washington, Seattle, WA, USA}

\author{Matthias Florian}
\affiliation{Department of Electrical and Computer Engineering, Department of Physics, University of Michigan, Ann Arbor, Michigan 48109, United States}

\author{Julian Klein}
\affiliation{Department of Materials Science and Engineering, Massachusetts Institute of Technology, Cambridge, Massachusetts 02139, USA}

\author{Kseniia Mosina}
\author{Zdenek Sofer}
\affiliation{Department of Inorganic Chemistry, University of Chemistry and Technology Prague, Technickaá 5, 166 28 Prague 6, Czech Republic}

\author{Xiaodong Xu}
\affiliation{Department of Physics, Department of Materials Science and Engineering, University of Washington, Seattle, WA, USA}

\author{Akashdeep Kamra}
\author{Francisco J. García-Vidal}
\affiliation{Departamento de Física Teórica de la Materia Condensada and Condensed Matter Physics Center (IFIMAC), Universidad Autónoma de Madrid, E- 28049 Madrid, Spain.}

\author{Andrea Al\`{u}}
\email{aalu@gc.cuny.edu}
\affiliation{Department of Physics, The Graduate Center, City University of New York, New York, NY 10016, USA}
\affiliation{Photonics Initiative, CUNY Advanced Science Research Center, New York, NY, 10031, USA}
\affiliation{Department of Electrical Engineering, City College of the City University of New York, New York, NY, 10031, USA}

\author{Vinod M. Menon}
\email{vmenon@ccny.cuny.edu}
\affiliation{Department of Physics, City College of New York, New York, NY 10031, USA}
\affiliation{Department of Physics, The Graduate Center, City University of New York, New York, NY 10016, USA}

\maketitle
\tableofcontents
\clearpage

%\nocite{Garcia2021,Schlawin2022,Bloch2022,Sentef2018,Ashida2020,Appugliese2022,Seyler2018,Zhang2019,Kang2020,Wilson2021,Klein2022-1,Huang2017,Wu2019,Hwangbo2021,Dirnberger2022,Canales2021,Klingshirn2012,Munkhbat2018,Fieramosca2018,Dang2020,Klein2022-2,Lim2017,Bajoni2008,Bae2022,Geoff2022,Cham2022,Lopez2022,Liu2022,Odonnel1991}

\section{Crystal structure}

\begin{figure*}[h!]
	\includegraphics[]{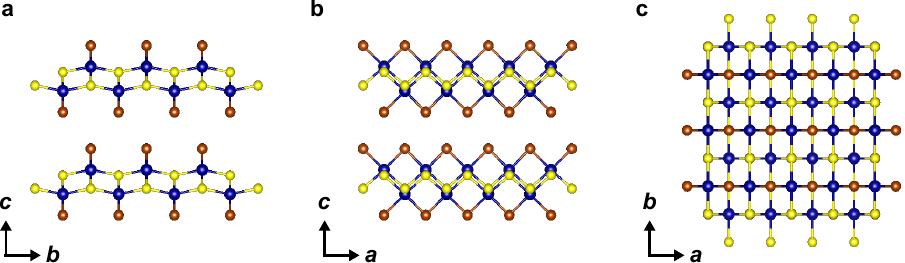}
	\caption{\textbf{Crystal structure of CrSBr. }
		\textbf{a}-\textbf{c}~Crystal structure of CrSBr viewed along the $a$--, $b$--, and $c$--direction, respectively. Cr atoms are depicted in blue, S atoms in yellow, and Br atoms in red. 
		\label{fig:SFig-Crystal_structure}}
\end{figure*}

%\clearpage

\section{Optical properties of mesoscopic CrSBr crystals: Theory and experiment}

\subsection{Exciton-polaritons in mesoscopic crystals}

To provide a better understanding of the complex optical properties of CrSBr that arise from strong coupling between excitons and photons, we use a semiclassical theory as well as a quantum-mechanical approach of the exciton-radiation interaction to gain access to optical spectra as well as the underlying exciton-polariton states.

\subsubsection{Dielectric function and transfer-matrix calculations}

The semiclassical theory of exciton-photon interaction is based on a susceptibility treatment of the optical response~\cite{claudio_andreani_optical_1995} within the CrSBr crystal. Maxwell's equations are solved including the frequency dependent material response using a tranfer-matrix formulation, from which linear optical spectra for different cavity structures are obtained.

The optical dielectric function for excitons can be derived microscopically using the semiconductor Bloch equations and written as~\cite{haug_quantum_2004}
\begin{align}
\varepsilon_{b}(\omega) = \varepsilon^\infty_b - \frac{\Delta_{X}}{\hbar\omega - \hbar\omega_X + i\gamma}
\label{Eq:elliot}
\end{align}
where $\hbar\omega_X$ and $\hbar/\gamma$ are the exciton energy and lifetime, respectively. Here, we consider only the lowest exciton level, while contributions due to all other resonances are included in a frequency-independent background relative permittivity $\varepsilon^\infty_b$. We note that the Elliot form~\eqref{Eq:elliot} only contains the resonant terms and that it is closely connected to a Lorentz oscillator model
\begin{align}
\varepsilon_{b}(E) = \varepsilon^\infty_b - \frac{f_{X}}{E^2 - E_X^2 + i\Gamma E}, 
\end{align}
by identifying $f_X = 2\sqrt{ E_X^2 - \gamma^2}\Delta_{X}$, $E_X^2 = \hbar\omega_X - \gamma^2$ and $\Gamma = 2\gamma$.

The exciton oscillator strength $\Delta_{X} = 2|d_b^{cv}|^2|\psi_X(\mathbf r=0)|^2$ is determined by the transition dipole element $d_b^{cv}$, which has predominant contribution along b-direction (c.f. \ref{fig:ExtData-6}). $\Psi(r)$ denotes the electron and hole relative wave function inducing Coulomb enhancement of the optical transition. Given the strong dielectric anisotropy, the tensorial character of the dielectric function needs to be taken into account. Bulk CrSBr crystallizes in an orthorhombic structure and we consider $\varepsilon_{a}(\omega) = \varepsilon^\infty_a$ and $\varepsilon_{c}(\omega) = \varepsilon^\infty_c$ for polarizations along a- and c-direction, respectively. 

\begin{figure*}[]
	\includegraphics[]{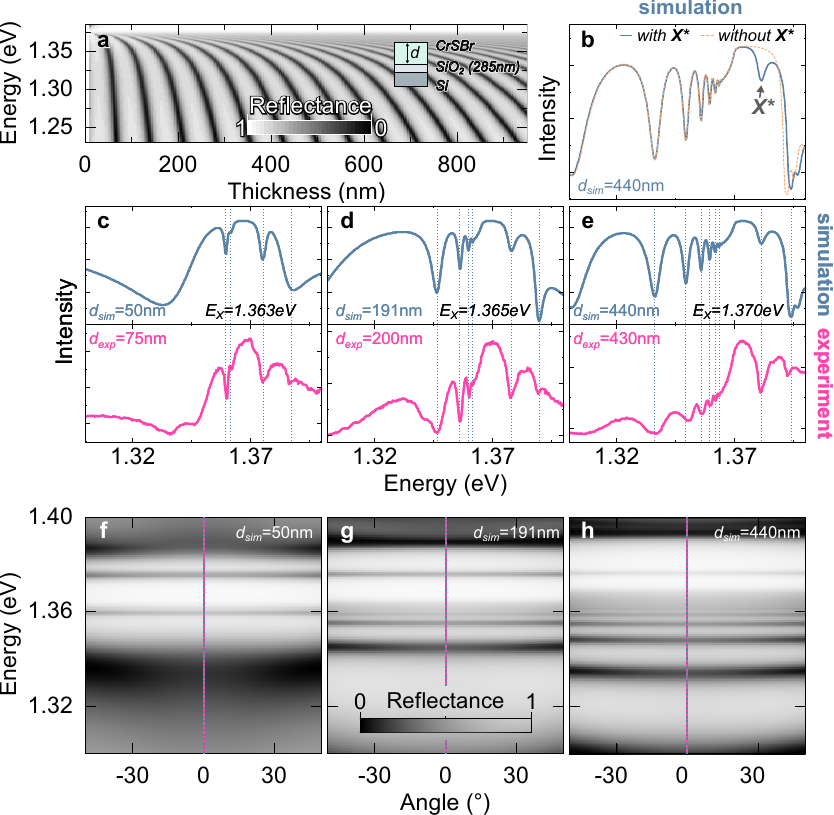}
	\caption{\textbf{Differential reflectance spectra of self-hybridized polaritons in CrSBr crystals on a $\textrm{SiO}_2/\textrm{Si}$ substrate.}	
		\textbf{a}~Polariton dispersion of CrSBr crystals on a $\textrm{SiO}_2$ (285\,nm)/$\textrm{Si}$ structure. 
		\textbf{b}~Calculated reflectance of a 440\,nm-thick flake with (solid line) and without (dotted line) a second, weak oscillator, $X^*$, at $\sim1.38$\,eV. Inclusion of $X^*$ has a negligible effect on the polariton states below 1.37\,eV analyzed in the main manuscript.
		\textbf{c}--\textbf{e}~Simulated and measured reflectance spectra of self-hybridized polaritons in three different CrSBr crystals. Crystal thickness $d_{sim}$ is extracted from the simulations, while $d_{exp}$ is determined by atomic force microscopy. $E_X$ indicates the energy of excitons used in the simulation.
		\textbf{f}--\textbf{h}~Simulated angle-resolved reflectance for polarization along the $b$--axis. 
		Optical signals were analyzed along the $b$--axis.
		\label{fig:SFig-exp-sim-CrSBr-on-SiO2}}
\end{figure*}

To obtain an estimate for the static dielectric tensor, spin-polarized density functional theory (DFT) calculations are carried out using \textsc{Quantum Espresso}~\cite{giannozzi_quantum_2009, giannozzi_advanced_2017} and considering an antiferromagnetic order along the stacking direction~\cite{Wilson2021}. We apply the generalized gradient approximation (GGA) by Perdew, Burke, and Ernzerhof (PBE) \cite{perdew_generalized_1996, perdew_generalized_1997} and use optimized norm-conserving Vanderbilt pseudopotential~\cite{van_setten_pseudodojo_2018} at a plane-wave cutoff of $80$~Ry. Uniform meshes with $8\times6\times2$ k-points are combined with a Fermi-Dirac smearing of $5$~mRy. We are using fixed lattice constant of a = $\SI{3.511}{\angstrom}$, b = $\SI{4.746}{\angstrom}$ and c = $\SI{7.916}{\angstrom}$ that are obtained from synchrotron XRD data of CrSBr~\cite{Lopez2022}. Structural relaxations were performed until the forces were smaller than $\SI{0.005}{\electronvolt\per\angstrom}$. The D3 Grimme method~\cite{GrimmeD3} is used to include van-der-Waals corrections. We find $\varepsilon_{a}=11.5$, $\varepsilon_{b}=43.1$, and $\varepsilon_{c}=9.1$.

\begin{figure*}[]
	\includegraphics[]{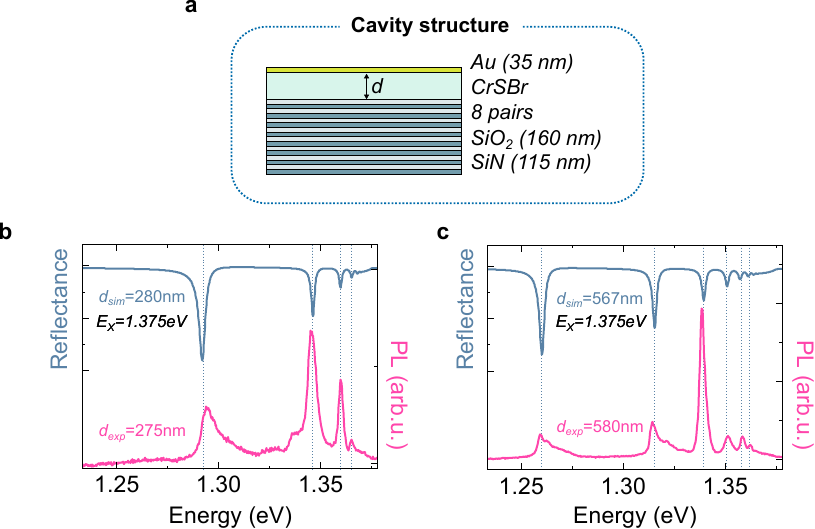}
	\caption{\textbf{Differential reflectance and PL emission of CrSBr crystals embedded between planar mirrors.}
		\textbf{a}~Layout of the CrSBr microcavity structure used in the simulations. 
		\textbf{b},\textbf{c}~Simulated reflectance and measured PL spectra of two CrSBr cavity samples.
		Crystal thickness $d_{sim}$ is extracted from the simulations, while $d_{exp}$ is determined by atomic force microscopy. $E_X$ indicates the energy of excitons used in the simulation. Optical signals were analyzed along the $b$--axis.
		\label{fig:SFig-exp-sim-CrSBr-cavity}}
\end{figure*}

Reflectance spectra of CrSBr crystals on a $\textrm{SiO}_2/\textrm{Si}$ substrate (\cref{fig:SFig-exp-sim-CrSBr-on-SiO2}a) as well as embedded between planar mirrors (\cref{fig:SFig-exp-sim-CrSBr-cavity}a) are calculated using a general transfer matrix formulation, suitable for the description of wave propagation through anisotropic media~\cite{rumpf_improved_2011}. Frequency dependent refractive indices for SiO$_2$~\cite{gao_refractive_2013}, Si~\cite{schinke_uncertainty_2015}, SiN~\cite{luke_broadband_2015} and Au~\cite{johnson_optical_1972} are used together with nominal layer thicknesses. The exciton optical dielectric function is systematically determined by comparing the polariton mode structure that is visible in the simulated and measured spectra (indicated by vertical dashed lines in \cref{fig:SFig-exp-sim-CrSBr-on-SiO2}c-e and \cref{fig:SFig-exp-sim-CrSBr-cavity}b-c) for different CrSBr crystals. 

Excellent agreement is obtained for all samples using $\Delta_X = \SI{908}{\milli\electronvolt}$, with a typical deviation of <5\% between the crystal thickness extracted by simulations and those determined by atomic force microscopy. Few meV shifts of the exciton transition energy are observed from sample to sample that we attribute to strain effects caused by the different thermal expansion of the materials surrounding the CrSBr crystals.

In CrSBr crystals on a $\textrm{SiO}_2/\textrm{Si}$ substrate an additional resonance is apparent at $\sim\SI{1.38}{\electronvolt}$, similar to the recently observed $X^*$ transition~\cite{Klein2022-1}. While it is barely visible in PL, the appearance in the differential reflectivity measurements suggests finite oscillator strength and band related transitions. It cannot be explained by upper polariton modes of the 1s exciton as shown in Fig.~\ref{fig:SFig-exp-sim-CrSBr-on-SiO2}b. Experimental reflectance spectra are well described by adding an additional $X^*$ resonance to Eq.~\eqref{Eq:elliot} with an energy splitting of the 1s exciton and the $X^{*}$ of $\Delta E \sim\SI{14}{\milli\electronvolt}$. Compared to the 1s exciton resonance, the $X^*$ has a significantly weaker oscillator strength of $\Delta_{X^*}\sim\SI{55}{\milli\electronvolt}$. As a result, the induced refractive index change has a negligible effect on the polariton states below 1.37\,eV that are analyzed in the main manuscript. 
The origin of $X^{*}$ is still unresolved.
It may involve momentum indirect transitions due to the strong extension of the wavefunction along the $\Gamma - X$ direction or transitions between the split conduction bands and the valence band~\cite{Klein2022-1}.

\begin{figure*}[]
	\includegraphics[width=\linewidth]{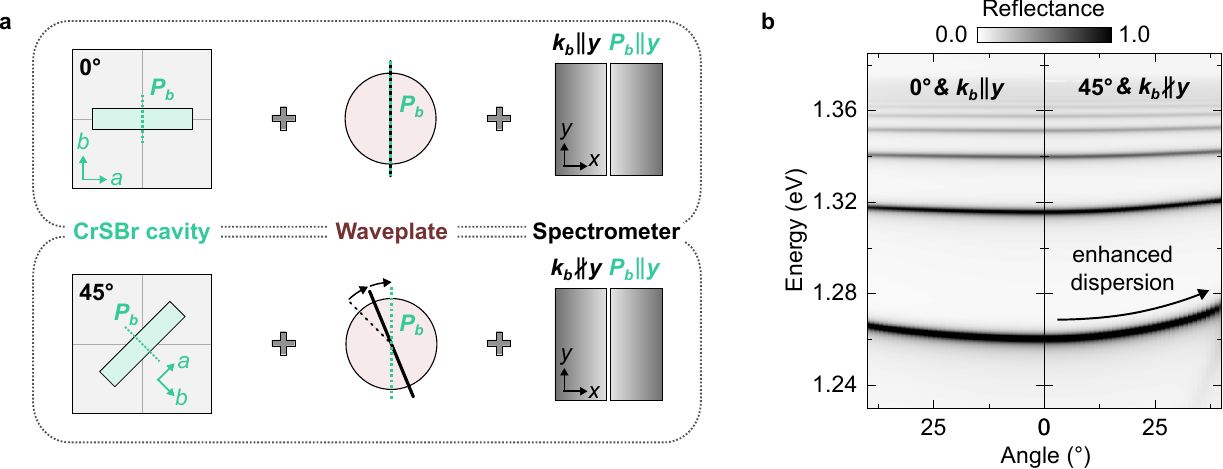}
	\caption{\textbf{Role of crystal orientation in the measured polariton dispersion}
		\textbf{a}~Upper panel: A CrSBr cavity crystal oriented with $b$--axis parallel to the slit of the spectrometer ($0^\circ$). Hence, the exciton polarization axis $\boldsymbol{P_b}$ coincides with the momentum direction $k_b$ imaged by the spectrometer. Lower panel: The same CrSBr cavity crystal rotated by $45^\circ$. A wave-plate rotates the exciton polarization axis $\boldsymbol{P_b}$ parallel to the slit, but the rotation of the $b$--axis with respect to the spectrometer changes the momentum direction imaged by the spectrometer. 
		\textbf{b}~Rotation of the sample images a stronger polariton dispersion, as demonstrated by the simulated reflectance map for $0^\circ$ and $45^\circ$ configurations.
		\label{fig:SFig-sim-0-vs-45-deg}}
\end{figure*}

The simulated polariton dispersion shown in Fig.~1D of the main text is determined by calculating reflectance spectra as a function of the elevation angle $\phi$, thereby changing the in-plane component of the incidence wave vector according to $k_\parallel = k_0\cos{\phi}$. Given the strong anisotropy of CrSBr, care has to be taken regarding the crystal orientation. If the CrSBr crystal is oriented with the $b$--axis parallel to the slit of the spectrometer ($0^\circ$) the exciton polarization axis $\boldsymbol{P_b}$ coincides with the momentum direction $k_b$ imaged by the spectrometer (c.f. Fig.~\ref{fig:SFig-sim-0-vs-45-deg}a), in which case the measured modes are TE polarized. By rotating the CrSBr cavity crystal by $45^\circ$ (azimuth) with respect to the plane-of-incidence (c.f. Fig.~\ref{fig:SFig-sim-0-vs-45-deg}b), the polariton modes get mixed TE and TM polarization, which results in a stronger polariton dispersion given that $\varepsilon_c < \varepsilon_b$, as shown in Fig.~\ref{fig:SFig-sim-0-vs-45-deg}c.

\subsubsection{Quantum model of strong light-matter coupling in mesoscopic crystals}

While a semiclassical theory is well suited to calculate optical spectra and polariton modes, a quantum mechanical approach is required to characterize the composition of the underlying polariton states. A quantum treatment of the exciton-photon interaction is based on a microsopic theory of the coupling between the excitons and the quantized electromagnetic modes. The approximation of keeping only quadratic terms yields a Hamiltonian which can be diagonalized exactly and the mixed exciton-photon modes are the polariton modes of the system.

We consider light polarized in b-direction and assume that the CrSBr crystal of thickness L is embedded in a microcavity with perfectly reflecting walls. The whole system keeps the translational invariance along the plane orthogonal to the z--direction. The dispersion of the electromagnetic modes is given by $\omega^c_{\mathbf q, q_z} = v \sqrt{\mathbf q^2 + q_z^2}$ with a z-dependence according to $\cos(q_z z)$. Here, $\mathbf q$ is the in-plane wave vector, $q_z = \pi n/L$ with integer values $n$ and $v=c/\sqrt{\varepsilon_b}$. 

The interaction between exciton and photon states can be calculated by representing the vector potential $\mathbf A(\mathbf r)$ in second-quantization~\cite{savona_quantum_1994}:
\begin{align}
\mathbf A(\mathrm r) = \sum_{\mathbf q, q_z} \sqrt{\frac{2\pi\hbar v}{\Omega (\mathbf q^2 + q^2_z)^{\frac{1}{2}}}} [b_{\mathbf q, q_z} e^{i\mathbf q\cdot\rho} - b^\dagger_{\mathbf q, q_z} e^{-i\mathbf q\cdot\rho}]\cos(q_z z) \mathbf e_{\mathbf q, q_z}
\end{align}
where $\Omega$ is the normalization volume of the photon eigenmodes and $\mathbf e_{\mathbf q, q_z}$ are the unit vectors of the transversal photon polarization.

Under weak excitation conditions and using the Coulomb gauge the exciton-photon interaction can be written in resonant approximation as~\cite{savona_quantum_1994}
\begin{align}
H_{\text{rad}} = -i\sum_{\mathbf q, k_z, q_z} C_{\mathbf q, q_z, k_z} (b^\dagger_{\mathbf q, q_z} X_{\mathbf q, k_z} - h.c.)\,.
\end{align}
In this expression, we define $X_{\mathbf q, k_z}$, $X^\dagger_{\mathbf q, k_z}$ as the annihilation and creation operator for an exciton with given in-plane wave vector $\mathbf q$ and subband index $k_z$. The exciton operator obeys Bose commutation relations. The interaction strength is given by
\begin{align}
C_{\mathbf q, q_z, k_z} = \frac{\omega^X_{\mathbf q, q_z}}{c} \sqrt{\frac{2\pi\hbar v}{L}} (q^2 + q^2_z)^{\frac{1}{2}} (\mathbf e_{\mathbf q, q_z} \cdot \mathbf d^{cv}_{\mathbf q, q_z}) \psi_X(\mathbf r=0) I_{k_z,q_z}
\end{align}
where $\omega^X_{\mathbf q, q_z}$ is the exciton dispersion and 
\begin{align}
I_{k_z,q_z} = \int^{L/2}_{-L/2} dz \rho_{k_z}^{cv}(z) e^{-i q_z z}
\end{align}
describes the overlap between the confinement function of electrons and holes in z-direction $\rho_{k_z}^{cv}(z)$ and the electromagnetic mode. In the bulk crystal limit we can assume that $I_{k_z,q_z}\approx \delta_{k_z,q_z}$.

A simplified expression for the exciton-photon coupling strength can be obtained assuming a weak $q$-dependence of the exciton dispersion and dipole matrix elements for relevant photon momenta:
\begin{align}
C_{\mathbf q, q_z} \approx g_{0} \sqrt{\frac{\omega_X}{\omega^c_{\mathbf q, q_z}}}
\label{Eq:coupling_strength}
\end{align}
with $g_{0} = \sqrt{\frac{\hbar\omega_X\Delta}{2\varepsilon_b}}$ and $\Delta$ the exciton oscillator strength. The bilinear exciton-photon Hamiltonian 
\begin{align}
H = \sum_{\mathbf q, q_z} \hbar\omega_X X^\dagger_{\mathbf q, q_z}X_{\mathbf q, q_z} + \sum_{\mathbf q, q_z} \hbar \omega^c_{\mathbf q, q_z} + H_{rad}
\end{align}
can be diagonalized by introducing polariton operators
\begin{align}
p_{\mathbf q,q_z} = u(\mathbf q,q_z)X_{\mathbf q,q_z} + v(\mathbf q,q_z)b_{\mathbf q,q_z}
\label{Eq:polariton_operator}
\end{align}
as a linear combination of exciton and photon operators~\cite{hopfield_theory_1958}. The idea is to demand the polariton operators to obey Bose commutation relations and the coefficients $u(\mathbf q,q_z)$ and $v(\mathbf q,q_z)$ are chosen so that the exciton-photon Hamiltonian becomes diagonal. As a result one obtains the polariton spectrum 
\begin{align}
\Omega^{UP/LP}_{\mathbf q,q_z} = \frac{1}{2}(\hbar\omega_X + \hbar \omega^c_{\mathbf q, q_z}) \pm  \frac{1}{2}\sqrt{(\hbar\omega_X - \hbar \omega^c_{\mathbf q, q_z})^2 + 4C_{\mathbf q, q_z}^2}
\label{Eq:polariton_dispersion}
\end{align}
showing upper (LP) and lower (LP) polariton branches. At the crossing point between exciton and cavity dispersion ($\hbar\omega_X = \hbar\omega^c_{\mathbf q, q_z}$), the polariton splitting is $2g_{0}$, which can be viewed as the vacuum-field Rabi splitting referred to in the main manuscript. Based on the material parameters of CrSBr obtained using the semiclassical approach, we can estimate a Rabi energy of $g_0 = ~\SI{120}{\milli\electronvolt}$. The exciton and photon fraction can be derived from the coefficients in Eq.\ref{Eq:polariton_operator} and noting that $|u_{\mathbf q,q_z}|^2 + |v_{\mathbf q,q_z}|^2 = 1$.

\begin{figure*}[]
	\includegraphics[]{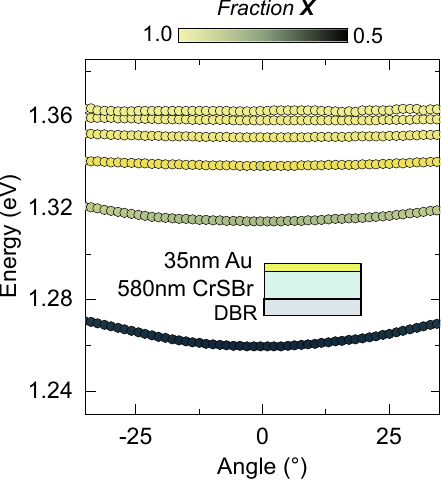}
	\caption{\textbf{Exciton-photon hybridization in CrSBr cavities.}
		Color-coded exciton fraction $\boldsymbol{X}$ plotted for the different polariton branches of our 580\,nm-thick cavity sample. 
		\label{fig:SFig-QuModel_fraction-X}}
\end{figure*}

In general, the dispersion relation obtained by a semiclassical and quantum mechanical approach can be seen to be equivalent, provided that the complex quantum-mechanical energies and the poles of the transmission coefficient coincide~\cite{savona_quantum_1995}. 
This problem has been solved for simple cavity structures close to the exciton frequency~\cite{savona_quantum_1995}, but presents a challenge for more complex cavity designs involving absorbing materials such as mesoscopic CrSBr crystals. 
We therefore use the experimentally measured polariton dispersion, which is in perfect agreement with our calculations based on the semiclassical model, and use~\eqref{Eq:coupling_strength} and \eqref{Eq:polariton_dispersion} to estimate $\omega^c_{\mathbf q, q_z}$ for each polariton mode (cf. \cref{fig:SFig-QuModel_fraction-X}).

\subsubsection{Comparison with other self-hybridized polariton systems in mesoscopic crystals}

Polaritons in CrSBr can be well-characterized by a dielectric function based on the parameters provided in Section S2. 
To compare the oscillator strength of excitons in CrSBr with those in other materials known to host self-hybridized polaritons, we reference values of the dielectric functions used to describe mesoscopic crystals of transition-metal dichalcogenides and perovskites. 

\begin{center}
	\begin{tabular}{ |c|c|c|c|c| } 
		\hline
		Material & $E_X$ (eV) & $\gamma_X$ (meV) & $f_x$ (eV$^2$) & Ref.\\
		\hline
		TMDC & 2.0 & 50 & 0.8 & \cite{Munkhbat2018} \\ 
		Perovskite & 2.4 & 50 & 1.1 & \cite{Fieramosca2018}  \\ 
		Perovskite & 2.4 & 30 & 0.9 & \cite{Dang2020} \\ 
		CrSBr & 1.37 & 2 & 2.5 & this work \\ 
		\hline
	\end{tabular}
\end{center}
\clearpage

\subsection{Comparison between three different experimental approaches to strong coupling in CrSBr}
\begin{figure*}[h!]
	\includegraphics[width=\linewidth]{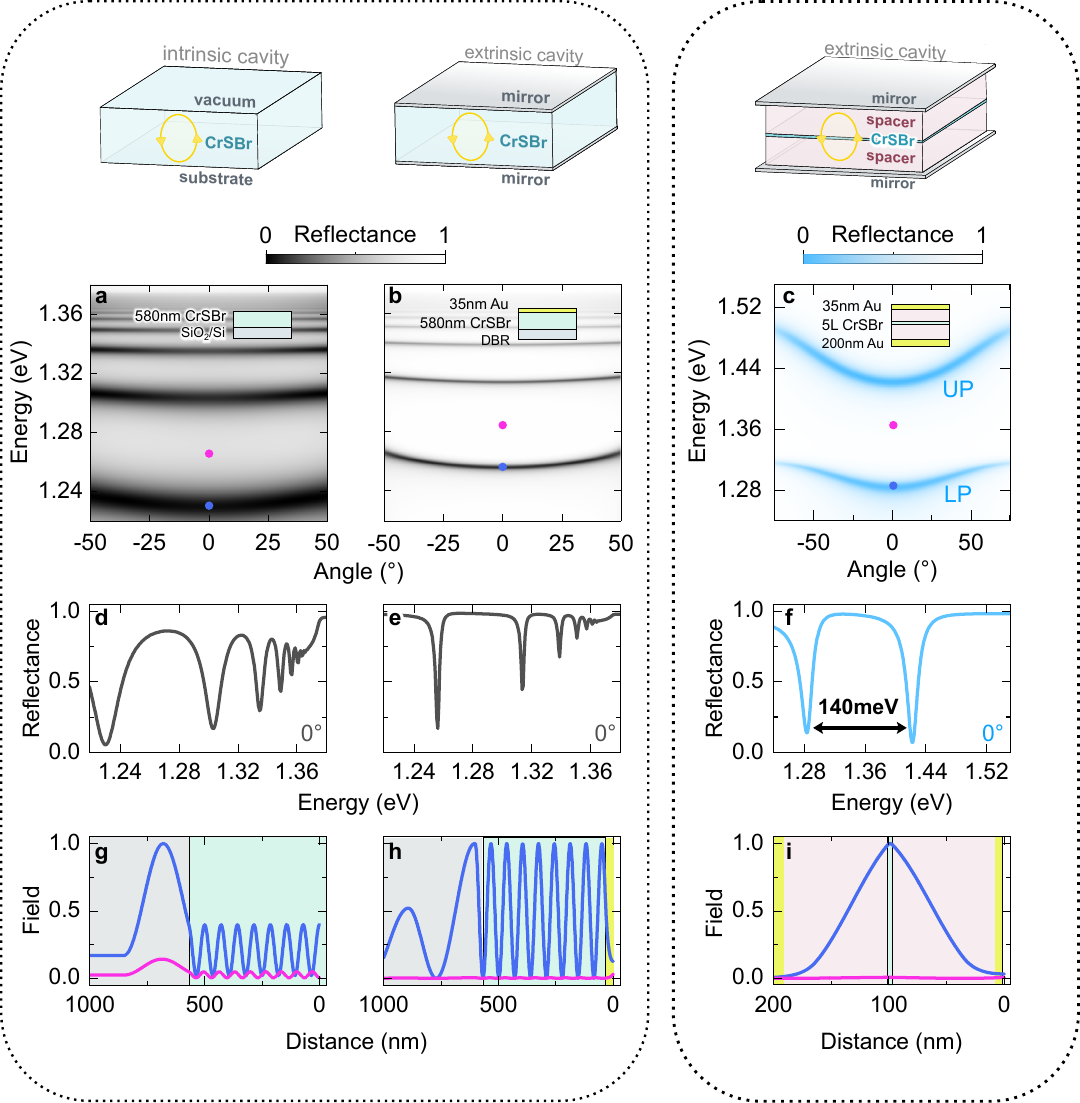}
\end{figure*}
\begin{figure*}[h!]
	\caption{\textbf{Polaritons in mesoscopic crystals and 2D polaritons in cavities.}
		\textbf{a} Simulated reflectance of a 580\,nm CrSBr flake on top of a $\textrm{SiO}_2/\textrm{Si}$ substrate with $\textrm{SiO}_2$ thickness of 285\,nm. 	
		\textbf{b}~Simulated reflectance of a 580\,nm crystal enclosed by a bottom dielectric Bragg mirror comprising of 8-pairs of alternating $\textrm{SiO}_2$/SiN layers and a 35\,nm-thin Au mirror. 
		\textbf{c}~Simulated reflectance of a 5L CrSBr flake enclosed by 62\,nm hexagonal boron nitride and 35\,nm and 200\,nm top and bottom Au layers.
		Reflectance spectra at normal incidence ($0^\circ$) show multiple sharp polariton modes with cavity quality factors around \textbf{d}~30 and \textbf{e}~300, and \textbf{f}~an upper polariton (UP) and a lower polariton (LP) branch with quality factor of 70. A normal-mode splitting of 140\,meV is observed in the case of 2D polaritons. 
		\textbf{g}--\textbf{i}~Respectively color-coded electric field profiles extracted for the energies indicated in the reflectance maps. Electric fields at the energy of polariton branches are strongly enhanced by the external cavity mirrors, while states at other energies are suppressed. In CrSBr crystals enclosed by external mirrors, the polariton field strength is enhanced by a factor of 300. For the bare crystal, the enhancement of polariton states over uncoupled states only amounts to a factor of 10. All reflectance plots are calculated for polarization along the $b$--axis.
		\label{fig:ExtData-1}}
\end{figure*}

\clearpage

\subsection{Emission and reflectance of crystals with and without external cavity mirrors}
A direct comparison of optical reflectance and emission spectra measured in bare flakes and in crystals enclosed by external cavity mirrors demonstrates an experimental advantage of adding external mirrors. 
Highly reflective cavity mirrors attached to the bottom and top surface of a mesoscopic crystal enhance the features of polariton states over any other, uncoupled states like those potentially arising from defects. 
Hence, in our samples that exhibit external cavity mirrors, we observe only emission peaks that perfectly match in energy with the resonances observed in reflectance, which in turn are in excellent agreement with our numerical calculation of the polariton spectrum. 
This correlation between states observed in emission and reflectance spectra is qualitatively different in bare crystals on $\textrm{SiO}_2/\textrm{Si}$.
While we clearly observe the signatures of polariton states in reflectance spectra, which match our numerical prediction, these signatures appear alongside other emission states in the PL spectra. 
Due to the weaker photon confinement in bare crystals compared to those enclosed by external cavity mirrors (see Fig. \cref{fig:ExtData-1}), polariton states are only moderately enhanced over other, uncoupled states. 
As a result, we observe many peaks that do not match the position of polaritons detected by reflectance spectroscopy. 

Overall, the complementary spectroscopic measurement of reflectance and emission spectra provides a straightforward approach to distinguish optical signatures of polaritons from other, still unidentified optical states in CrSBr, if the approach is complemented by accurate numerical prediction of the spectral positions of polaritons using the transfer-matrix method and the parameters given in Section S2. 

\begin{figure*}[h!]
	\includegraphics[]{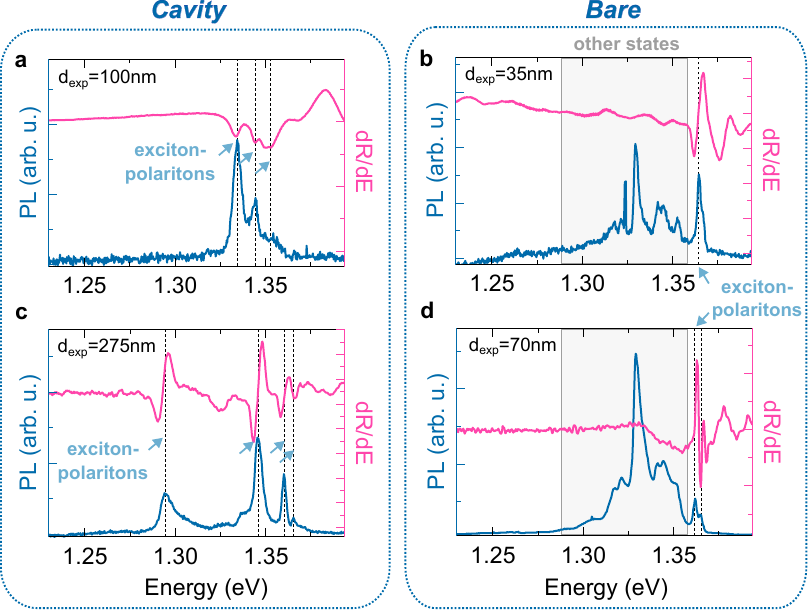}
	\caption{\textbf{Low-temperature PL emission and differential reflectance contrast dR/dE.}
		\textbf{a},\textbf{c}~In CrSBr crystals enclosed by external cavity mirrors, optical states observed in reflectance and PL perfectly match the theoretically predicted exciton-polariton dispersion.
		\textbf{b},\textbf{d}~In bare CrSBr crystals on $\textrm{SiO}_2/\textrm{Si}$ substrates, reflectance measurements match the predicted exciton-polariton dispersion, but other, unidentified states are observed in the low-temperature PL response at energies below the exciton-polaritons, highlighting the benefit of adding external cavity mirrors. Optical signals were analyzed along the $b$--axis.
		\label{fig:ExtData-2}}
\end{figure*}
\clearpage

\subsection{Polarization-dependent PL emission and reflectance of the 580\,nm cavity sample}
\begin{figure*}[h!]
	\includegraphics[]{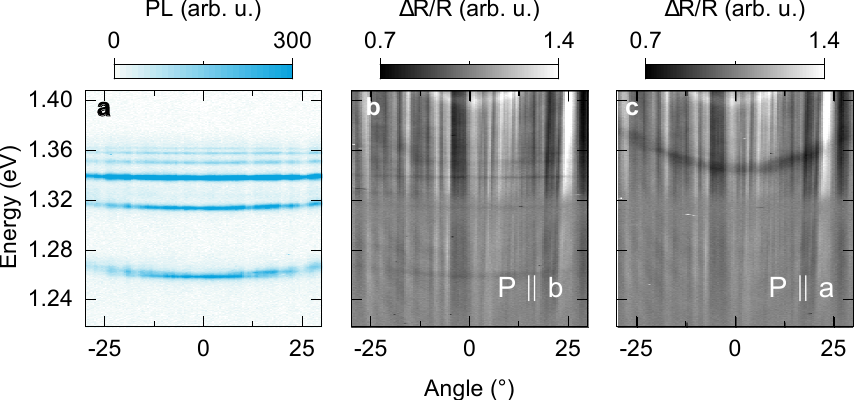}
	\caption{\textbf{Polariton emission and reflectance of the 580\,nm cavity sample.}
		\textbf{a}~Angle-resolved polariton PL emission recorded without polarization optics in the detection path.
		\textbf{b}~Angle-resolved differential reflectance analyzed for polarization $P$ along the in-plane $b$--axis shows exciton-polariton modes. 
		\textbf{c}~Angle-resolved differential reflectance analyzed for polarization $P$ along the in-plane $a$--axis shows a purely photonic cavity mode.  
		Data recorded at 1.6\,K. 
		\label{fig:ExtData-6}}
\end{figure*}

\Cref{fig:ExtData-6} compares the low-temperature PL emission and differential reflectance of our 580\,nm cavity sample for polarization analyzed along the $b$--axis, showing perfect agreement between the optical states observed in both experiments.  
When we analyze polarization along the $a$-axis, we observe no PL signal and an optical cavity mode at $\sim1.3e$\,eV that does not couple to the polarization of the excitons oriented along the $b$--axis. 
This optical cavity mode is therefore completely independent of the dispersion of excitons and polaritons and is only subject to the usual thickness dependence of optical cavity modes.  

\clearpage

\subsection{Field-induced modification of the polariton effective mass}
\begin{figure*}[h!]
	\includegraphics[width=0.7\linewidth]{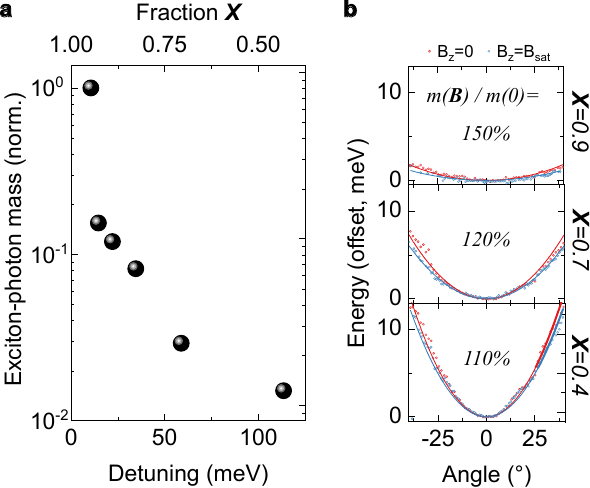}
	\caption{\textbf{Magnetic field dependent polariton effective mass in the 580\,nm cavity sample.}
		\textbf{a}~Relative changes of the polariton effective mass for different branches with zero-angle exciton fraction $\boldsymbol{X}$ determined from angle-resolved PL emission. 
		\textbf{b}~Application of a saturation field induces a detuning-dependent change of the polariton effective mass. Relative changes in the effective polariton mass are found to vary for different zero-angle exciton fractions. 
		Effective mass is determined from the curvature of parabolic fits. Data was collected at 1.6\,K and analyzed along the $b$--axis. Red (blue) dots and lines represent data and fits obtained at $B_z=0$ ($B_z=B_{sat}$). 	
		\label{fig:SFig-Effective Mass}}
\end{figure*}
\clearpage

%%%%%%%%%%%%%%%%%%%%%%%%%%%%%%%%%%%%%%%%%%%%%%%%%%%%%%%%%%%%

%%%%%%%%%%%%%%%%%%%%%%%%%%%%%%%%%%%%%%%%%%%%%%%%%%%%%%%%%%%%

%%%%%%%%%%%%%%%%%%%%%%%%%%%%%%%%%%%%%%%%%%%%%%%%%%%%%%%%%%%%

\section{Theoretical analysis of the effects of coherent and incoherent magnons} \label{Section: Theoretical analysis of the impact of coherent and incoherent magnons}

Here, we examine the influence of the magnetic order and its fluctuations on exciton and exciton-polariton energies in CrSBr. 
Specifically, our analysis considers effects on the exciton energy produced by external magnetic fields and optically induced coherent magnons in the presence of spin canting at low temperatures. 
Moreover, we evaluate the role of non-zero temperatures both in a model based on incoherent magnons, as well as by including temperature effects in a description of sub-lattice magnetizations. 
We thus derive an analytic model for the temperature- and field-dependence of the exciton energy in CrSBr. 

\subsection{Coupling of magnetic and electronic structure at zero temperature}

We first analyze the relation between the electronic and magnetic structure of CrSBr at zero temperature in the presence of an out-of-plane (spin-canting) external magnetic field, analogous to the discussion presented in Ref.~\cite{Wilson2021}. 
This will serve as a framework for our further analysis detailed below.

\subsubsection{Magnetic ground state with applied magnetic field}

We consider the antiferromagnet to be described by the magnetic free energy density:
\begin{align}\label{eq:free}
F & = J \pmb{M}_A \cdot \pmb{M}_B + K_h \left( M_{Ac}^2 + M_{Bc}^2\right) - K_e \left( M_{Ab}^2 + M_{Bb}^2\right) - \mu_0 H_0 \left( M_{Ac} + M_{Bc} \right),
\end{align}
where $\pmb{M}_{A,B}$ are the magnetizations, assumed spatially uniform, of sub-lattices A and B. In Eq.~\eqref{eq:free} above, $J$ ($>0$) parameterizes the antiferromagnetic exchange between the two sub-lattices, $K_h$ ($>0$) accounts for the hard-axis anisotropy in the out-of-plane $c$--direction, $K_e$ ($>0$) captures the easy-axis anisotropy along the $b$--axis, and the final term on the right represents Zeeman energy due to externally applied field $\pmb{H}_{\mathrm{ext}} = H_0 \hat{\pmb{c}}$.

\begin{figure}[]
	\centering
	\includegraphics[]{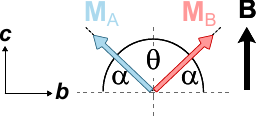}
	\caption{Schematic depiction of the two sub-lattice magnetizations in equilibrium. We assume an easy axis along $\hat{\pmb{b}}$ and a hard axis along $\hat{\pmb{c}}$. An external magnetic field is applied along $\hat{\pmb{c}}$.}\label{fig:coh}
\end{figure}

Parameterizing the equilibrium configuration as depicted in Fig.~\ref{fig:coh} via the angle $\alpha$, the free energy density simplifies to:
\begin{align}
F & = - J M_0^2 \cos 2\alpha + 2 K M_0^2 \sin^2 \alpha - 2 \mu_0 H_0 M_0 \sin \alpha - 2 K_e M_0^2,
\end{align}
where $M_0$ is the magnitude of sub-lattice magnetizations $\pmb{M}_{A,B}$, and we define $K \equiv K_e + K_h$. The equilibrium configuration is determined by minimizing the free energy:
\begin{align}
\frac{\partial F}{\partial \alpha} & = 0, \\
\implies \sin \alpha & = \frac{\mu_0 H_0}{2 M_0 \left( J + K\right)}, \quad \frac{\mu_0 H_0}{2 M_0 \left( J + K\right)} < 1  \nonumber \\
\alpha & = \pi/2, \quad \frac{\mu_0 H_0}{2 M_0 \left( J + K\right)} \geq 1. \label{eq:alpha}
\end{align}

\subsubsection{Exciton energy shift in a spin-canting external field}

Considering the role of interlayer tunneling and its spin dependence, the exciton energy shift may be expressed as~\cite{Wilson2021}:
\begin{align}
\Delta E & = \Delta_B \cos^2 \left( \frac{\theta}{2} \right) = \Delta_B \cos^2 \left( \frac{\pi}{2} - \alpha \right), \\ 
& = \Delta_B \sin^2 \left(\alpha\right),
\end{align}
where $\theta$ is the angle between $\pmb{M}_A$ and $\pmb{M}_B$ (Fig.~\ref{fig:coh}), $\Delta_B$ is the maximum field-induced shift of the exciton energy, and $\alpha$ is given by Eq.~\eqref{eq:alpha}. Thus, we see that the exciton energy shift varies quadratically with the applied magnetic field until a ferromagnetic configuration is realized, after which the exciton energy remains constant (cf. bell-like curve in Fig.~2A of the main manuscript).

\subsubsection{Exciton energy shift in the presence of coherent magnons}

A recent pump-probe experiment demonstrated the dynamic aspects of this coupling between the magnetic and electronic structure in CrSBr using coherent magnons induced by ultrashort optical excitation~\cite{Bae2022}. 
In a canted configuration, where $\theta_{0}$ is the equilibrium angle between the two sub-lattice magnetizations governed by a static external field, coherent magnons produce a time-dependence of the angle $\theta$
\begin{align}
\theta (t) = \theta_0 + \Delta \theta_0 \exp(-\kappa t) \cos(\omega t)\,, 
\end{align}
where $\kappa$ is the magnon decay rate and $\omega$ is their frequency. 

If we assume that $\Delta \theta_0 \ll \theta_0$, we can express the effect of coherent magnons on the exciton energy by
\begin{align}
\Delta E &= \Delta E (\theta = \theta_0) + \eval{\dv{}{\theta}\Delta E}_{\theta = \theta_0}  (\Delta \theta_0 \exp(-\kappa t)\cos(\omega t)) \label{Eq:Delta-i} \\
\dv{\theta} \Delta E &= \Delta_B \cos\left( \frac{\theta}{2} \right) \left(-\sin\left( \frac{\theta}{2} \right)\right) \nonumber \\
& = - \frac{\Delta_B}{2} \sin(\theta) \nonumber \\
\eval{\dv{\theta} \Delta E}_{\theta = \theta_0} & = - \frac{\Delta_B}{2} \sin(\theta_0)  \label{Eq:Delta-ii}
\end{align}

By inserting \cref{Eq:Delta-ii} into \cref{Eq:Delta-i}, we thus obtain an expression for the energy of excitons in the presence of spin canting and coherent magnons, 
\begin{align}
\Delta E (t) = \Delta_B \cos^2 \left( \frac{\theta_0}{2} \right) - \frac{\Delta_B}{2} \, \sin(\theta_0) \, \Delta \theta_0 \, \exp(-\kappa t) \, \cos(\omega t) \label{Eq:ExCohMag}\, .
\end{align}
This expression is used to numerically simulate the response of excitons-polaritons to coherent magnons in Fig.~3B of the main manuscript (cf.~also \cref{SSec:NumModel-exc-coh-magnon}).

\subsection{Coupling of magnetic and electronic structure at  low temperature}

\begin{figure}[]
	\centering
	\includegraphics[]{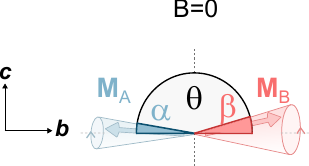}
	\caption{Schematic depiction of the two sub-lattice magnetizations at a moment of time. No external magnetic field is applied now and thus, the equilibrium configuration corresponds to $\alpha = \beta = 0$. At a given instant of time, $\alpha$ and $\beta$ have random values due to thermal fluctuations.}\label{fig:incoh}
\end{figure}

To address the effect of finite temperatures, we consider the same magnet and anisotropies as in the previous section [Eq.~\eqref{eq:free}], but first derive a description in the absence of any applied magnetic field, i.e., $H_0 = 0$. Thus, at zero temperature, the ground state corresponds to antiparallel sub-lattice magnetizations oriented along the $b$--axis. However, when one considers non-zero temperatures, thermal fluctuations cause the magnetizations to fluctuate about their equilibrium positions. At temperatures much smaller than the N\'eel temperature, these fluctuations are synonymous with the presence of thermally generated incoherent magnons in the system. An instantaneous configuration of the magnet is depicted in Fig.~\ref{fig:incoh}, where $\alpha$, $\beta$, and $\theta$ become random variables. We now examine the impact of such fluctuations, representing the presence of incoherent magnons in the system, on exciton energies. 

\subsubsection{Exciton energy shift in the presence of magnetic fluctuations}

Following the previous section, we may write the instantaneous exciton energy assuming that the electrons adapt instantly to the relatively slow magnetic fluctuations:
\begin{align}
\Delta E & = \Delta_B \cos^2 \left( \frac{\theta}{2} \right) = \Delta_B \cos^2 \left( \frac{\pi}{2} - \frac{\alpha + \beta}{2} \right), \\ 
& = \Delta_B \sin^2 \left( \frac{\alpha + \beta}{2} \right), \\
& = \Delta_B \left[ \sin \left( \frac{\alpha}{2} \right) \cos \left( \frac{\beta}{2} \right) + \sin \left( \frac{\beta}{2}  \right) \cos \left( \frac{\alpha}{2} \right)  \right]^2 , \\
& = \Delta_B \left[ \sin^2 \left( \frac{\alpha}{2} \right) \cos^2 \left( \frac{\beta}{2} \right) + \sin^2 \left( \frac{\beta}{2}  \right) \cos^2 \left( \frac{\alpha}{2} \right) + \frac{1}{2} \sin \alpha \sin \beta \right].
\end{align}
The experimentally recorded shift in the exciton energy is obtained by averaging, denoted by the operator $\avg{\cdot}$, over the thermal fluctuations:
\begin{align}
\avg{\Delta E} & = \Delta_B \left[ \avg{\sin^2 \left( \frac{\alpha}{2} \right)} \avg{ \cos^2 \left( \frac{\beta}{2} \right)} + \avg{\sin^2 \left( \frac{\beta}{2}  \right)} \avg{ \cos^2 \left( \frac{\alpha}{2} \right)} \right], \label{eq:eshift}
\end{align}
where we have assumed that the thermal fluctuations in sub-lattices A and B behave independently, and $\avg{\sin\alpha} = \avg{\sin\beta} = 0$. The latter assumption is well-justified since the average value of $\alpha$ and $\beta$ is 0. The former assumption essentially disregards the correlations that are generated by the antiferromagnetic exchange. This assumption is justified at temperatures larger than a few Kelvins since the antiferromagnetic exchange in the material under investigation has been found to be small (less than 1\,T)~\cite{Cham2022}, and can be neglected when we consider thermal fluctuations at finite temperatures.

\subsubsection{Exciton energy shift in the presence of incoherent magnons}
It is much more convenient and powerful to describe these incoherent fluctuations in terms of magnons, rather than the classical Landau-Lifshitz description that we have followed thus far. 
We now bridge the two pictures and relate the exciton energy with the incoherent magnon density. 
To this end, we note that the average projection of the sub-lattice magnetizations along the equilibrium $b$--direction, $M_{Ab}$ and $M_{Bb}$, are reduced by the existence of magnons:
\begin{align}
M_{Ab} & = M_0 \cos \alpha , \\
M_{Ab}^2 & = M_0^2 \cos^2 \alpha = M_0^2 (1 - \sin^2 \alpha), \\
\avg{M_{Ab}} & \approx M_0 - \frac{M_{0}}{2} \avg{\sin^2 \alpha}.
\end{align}
Equating this reduction in the magnetization to what is caused by magnons, we obtain:
\begin{align}
\frac{M_{0}}{2} \avg{\sin^2 \alpha} & = n_{A} \hbar \gamma, \\
\implies \avg{\sin^2 \alpha} & = \frac{2 n_{A} \hbar \gamma}{M_0},
\end{align}
where we have assumed the sub-lattice A magnons to bear spin $\hbar$, $\gamma$ ($>0$) is the sub-lattice gyromagnetic ratio, and $n_{A}$ is the density of magnons with spin opposite to the A sub-lattice equilibrium spin. Similarly, we obtain:
\begin{align}
\avg{\sin^2 \beta} & = \frac{2 n_{B} \hbar \gamma}{M_0},
\end{align}
for the density of magnons with spin opposite to the B sub-lattice equilibrium spin.

Substituting these in Eq.~\eqref{eq:eshift}, we obtain our final result:
\begin{align}
\avg{\Delta E} & = \Delta_B \left[ \avg{\sin^2 \left( \frac{\alpha}{2} \right)} \avg{ \cos^2 \left( \frac{\beta}{2} \right)} + \avg{\sin^2 \left( \frac{\beta}{2}  \right)} \avg{ \cos^2 \left( \frac{\alpha}{2} \right)} \right], \\
& = \Delta_B \left[ \avg{\sin^2 \left( \frac{\alpha}{2} \right)}  + \avg{\sin^2 \left( \frac{\beta}{2}  \right)} - 2 \avg{ \sin^2 \left( \frac{\alpha}{2} \right)} \avg{ \sin^2 \left( \frac{\beta}{2} \right)} \right], \\
& \approx \Delta_B \left[ \avg{\sin^2 \left( \frac{\alpha}{2} \right)}  + \avg{\sin^2 \left( \frac{\beta}{2}  \right)} \right], \\
& = \Delta_B \frac{\left(n_{A} + n_{B}\right) \hbar \gamma}{2 M_0}, \label{eq:dele}
\end{align}
where we have employed $\sin \left(\alpha/2 \right) \approx \left(\sin \alpha \right)/2$ and similar for $\beta$, valid when $\alpha, \beta \ll 1$.

According to Eq.~\eqref{eq:dele}, the shift of the exciton energy should vary linearly with the magnon temperature, since $n_{A,B} \sim k_B T$. However, this is not the complete story because our analysis assumes $M_0$ and other parameters entering the free energy to be temperature independent. When one takes this into account, for example by considering that $M_0$ is decreasing as the temperature increases~\cite{Cham2022}, the energy shift above bears a quadratic contribution in addition to the linear one. We address this regime by accounting for the reduction of the sublattice magnetizations below.

\subsubsection{Exciton energy shift in the presence of incoherent magnons and a saturation magnetic field}

In the previous section, we have considered zero external magnetic field. If we consider a large applied field, the ground state of the system is in a FM configuration, i.e., the two sub-lattice magnetizations align along the external magnetic field. We now consider the exciton energy shift under such a configuration.

\begin{figure}[]
	\centering
	\includegraphics[]{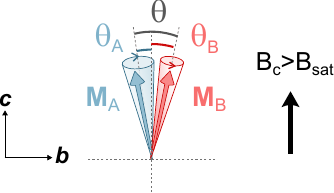}
	\caption{Schematic depiction of the two sub-lattice magnetizations at a moment of time. A large external magnetic field is assumed such that the equilibrium configuration corresponds to $\pmb{M}_A = \pmb{M}_B = M_0 \hat{\pmb{b}}$ [Eq.~\eqref{eq:alpha}]. $\theta_A$ and $\theta_B$, with $\theta = \theta_A + \theta_B$, now become the random variables with zero mean value associated with the thermal fluctuations.}\label{fig:incoh_highfield}
\end{figure}

Depicting the instantaneous configuration in Fig.~\ref{fig:incoh_highfield}, the exciton energy shift is evaluated as:
\begin{align}
\Delta E & = \Delta_B \cos^2 \left( \frac{\theta}{2} \right) = \Delta_B \cos^2 \left( \frac{\theta_A + \theta_B}{2} \right), \\ 
& = \Delta_B \left[ \cos \left( \frac{\theta_A}{2} \right) \cos \left( \frac{\theta_B}{2} \right) - \sin \left( \frac{\theta_A}{2}  \right) \sin \left( \frac{\theta_B}{2} \right)  \right]^2 , \\
& = \Delta_B \left[ \cos^2 \left( \frac{\theta_A}{2} \right) \cos^2 \left( \frac{\theta_B}{2} \right) + \sin^2 \left( \frac{\theta_A}{2}  \right) \sin^2 \left( \frac{\theta_B}{2} \right) - \frac{1}{2} \sin \theta_A \sin \theta_B  \right].
\end{align}
A shift equivalent to that recorded in our time-integrated optical experiments presented in Fig.~3 of the main manuscript is obtained by averaging over thermal fluctuations:
\begin{align}
\avg{\Delta E} & = \Delta_B \left[ \avg{\cos^2 \left( \frac{\theta_A}{2} \right)} \avg{ \cos^2 \left( \frac{\theta_B}{2} \right)} + \avg{\sin^2 \left( \frac{\theta_A}{2}  \right)} \avg{ \sin^2 \left( \frac{\theta_B}{2} \right)}  \right], \\
& =  \Delta_B \left[ 1 - \avg{\sin^2 \left( \frac{\theta_A}{2} \right)} - \avg{ \sin^2 \left( \frac{\theta_B}{2} \right)} + 2 \avg{\sin^2 \left( \frac{\theta_A}{2}  \right)} \avg{ \sin^2 \left( \frac{\theta_B}{2} \right)}  \right], \nonumber \\
& \approx \Delta_B \left[ 1 - \avg{\sin^2 \left( \frac{\theta_A}{2} \right)} - \avg{ \sin^2 \left( \frac{\theta_B}{2} \right)} \right].
\end{align}
Employing the analysis similar to that in the previous section, we may express this in terms of the magnon densities as follows
\begin{align}
\avg{\Delta E} & = \Delta_B \left(  1 - \frac{\left(n_{A} + n_{B}\right) \hbar \gamma}{2 M_0} \right). \label{eq:dele2}
\end{align}
Hence, we note that the exciton energy shift due to a non-zero population of incoherent magnons in this case is in the opposite direction as compared to the previous case Eq.~\eqref{eq:dele}. This makes sense as in the current configuration, the energy shift is already at maximum ($\Delta E=\Delta_B$) at zero temperature and the thermal fluctuations can only reduce the effect. 
%On the other hand, the main effect of the high magnetic field in this case is to reduce the magnon density $n_{A,B} \sim \exp(- \mu_0 H_0 \mu_B / k_B T)$, where $\mu_B$ is the Bohr magneton. Thus, at such large fields, the thermal fluctuation effects is much weaker, in consistence with the experiments.

\paragraph*{Discussion}

The analysis above has employed the hierarchy of energy scales (temperature much larger than antiferromagnetic exchange and anisotropies) present in the system and provided a simplified way to estimate the effects. A more rigorous analysis of the magneto-electric effect due to incoherent magnons, especially at low temperatures, should consider the fact that antiferromagnetic magnons do not always bear spin 1, when accounting for anisotropies that are important at low energies~\cite{Kamra2017,Kamra2020}. 
In fact, for the dominant easy-plane anisotropy in the material under investigation, the low-energy magnons bear a much smaller spin than 1. Such effects have been disregarded in our analysis and are expected to weaken the linear-in-temperature exciton energy shift at low temperatures. 

\subsubsection{Exciton energy shift induced by changes in sub-lattice magnetizations for a broader temperature range}

So far, we have used the expression for the exciton energy shift 
\begin{align}
\Delta E = \Delta_B \cos^2 \left( \frac{\theta}{2} \right) \, ,
\end{align}
which holds only as long as temperature-induced changes in the sub-lattice magnetizations are small. 
To account for the temperature dependence of the sublattice magnetization over a broad temperature range up to the ordering temperature, we express the exciton energy shift in terms of the magnetization: 
\begin{align}
\Delta E^\prime &= k\, \pmb{M}_A\cdot\pmb{M}_B \nonumber\\
&= k\, {M_S}^2 \cos \theta \, ,
\end{align}
where $M_S \equiv M_S(T)$ is the temperature-dependent magnetization of each sub-lattice and $k$ is a constant. 

Since they are based on the same underlying phenomena, the two different expressions for the exciton energy shift, $\Delta E$ and $\Delta E^\prime$, should be consistent. 
\begin{align}
\Delta E^\prime &= k\, {M_S}^2 \cos \theta \nonumber\\
&= k\, {M_S}^2 \left(2 \cos^2 \frac{\theta}{2}-1 \right) \nonumber\\
&= 2 k\, {M_S}^2 \cos^2 \frac{\theta}{2} - k\, {M_S}^2 \,
\end{align}

Using $\Delta_B = 2 k\, {M_S}^2$, we obtain
\begin{align}
\Delta E^\prime &= \Delta_B \cos^2 \left( \frac{\theta}{2} \right)  - k\, {M_S}^2 \nonumber\\
\Delta E^\prime &= \Delta E - k\, {M_S}^2 \nonumber\, ,
\end{align}
and thus $\Delta E$ and $\Delta E^\prime$ are fully consistent, but as we show in the following, $\Delta E^\prime$ is much more convenient to describe the temperature dependence of exciton energies in CrSBr. 
To account for the second-order phase transition, we assume a temperature-dependence of the sub-lattice magnetization
\begin{align}
\frac{M_S}{M_{0}} = \left(1-\frac{T}{T_{corr}}\right)^\beta \,
\end{align}
where $M_{0}$ is the magnetization at $T=0$, and $T_{corr}$ is the temperature up to which short-range correlations are observed (see \cref{SSec:NumModel-exc-coh-magnon}). 
As a result, the temperature- and field-dependence of the exciton energy in CrSBr is given by
\begin{align}
\Delta E^\prime &= k\, {M_{0}}^2 \cos(\theta) \left(1-\frac{T}{T_{corr}}\right)^{2\beta} \, ,
\end{align}
and explicit expressions for the temperature-dependence of the exciton energy for AFM ($\theta=\pi$) and FM ($\theta=0$) configurations are
\begin{align}
\Delta E^\prime &= - \,\frac{\Delta_B}{2} \left(1-\frac{T}{T_{corr}}\right)^{2\beta} & \text{for AFM}\, , \label{Eq:IncMagAFM}\\
\Delta E^\prime &= + \, \frac{\Delta_B}{2} \left(1-\frac{T}{T_{corr}}\right)^{2\beta} & \text{for FM} \label{Eq:IncMagFM} \, ,
\end{align}
For $T\ll T_{corr}$, we thus recover the linear temperature-dependence expected from the above analysis of incoherent magnons in the low temperature limit:
\begin{align}
\Delta E^\prime = \frac{\Delta_B}{2}\, \cos\theta_{0}\, \left( 1-\frac{2\beta T}{T_{corr}} \right) \, . 
\end{align}
In this analysis, $\Delta_B=-17.5$\,meV, as determined by numerical simulation of reflectance spectra. 
\clearpage

\section{Experimental and numerical analysis of polariton-magnon coupling}

\subsection{Measurement of the coupling between polaritons and coherent magnons}
\begin{figure*}[h!]
	\includegraphics[]{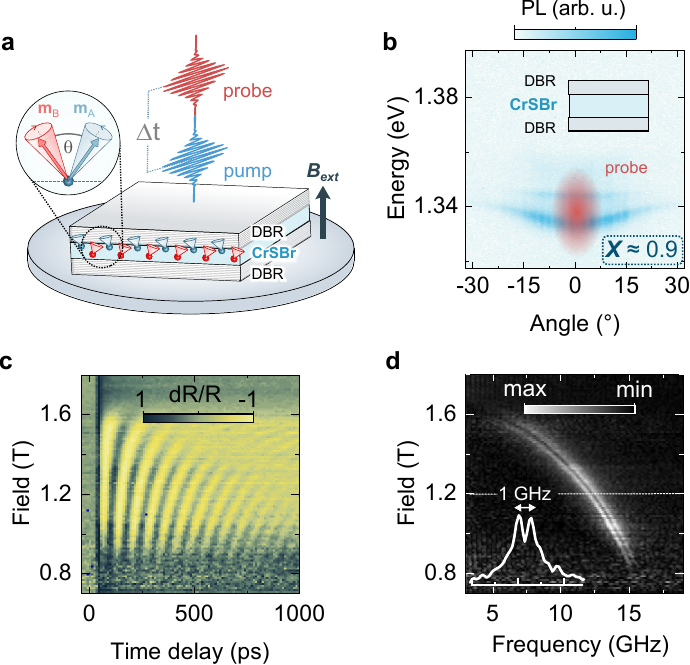}
	\caption{\textbf{Polariton-magnon coupling dynamics probed by resonant transient reflectance.}
	\textbf{a}~Schematic illustrating the transient reflectivity measurement employing a pump pulse with above-gap energy and a resonantly tuned probe pulse with a time delay $\Delta t$. 
	\textbf{b}~Corresponding polariton dispersion measured at 1.6K analyzed along the $b$--axis by polariton PL emission showing two closely-spaced polariton branches with exciton fractions $\boldsymbol{X}\approx0.9$. Red-shaded area indicates the spectrum of the probe pulse. 
	\textbf{c}~Pump-probe signal for different magnetic fields ($\boldsymbol{B_{ext}}\parallel c$). 
	\textbf{d}~Fourier transform of the pump-probe dynamics obtained by scanning the magnetic-field. Inset: Fourier spectrum obtained at $\boldsymbol{B_{ext}}=0.9$\,T showing a $\sim$1\,GHz frequency splitting attributed in a recent report to the coupling between acoustic phonons and magnons~\cite{Bae2022}.
	\label{fig:SFig-Experiment_pump-probe_B-field_dependence}}
\end{figure*}

\subsection{Numerical analysis of polariton-magnon coupling: Coherent magnons} \label{SSec:NumModel-exc-coh-magnon}

Our experimental results presented in Fig.~2B of the main manuscript allow us to model the time-dependent response of exciton-polaritons to a density of coherent magnons. 
We modify \cref{Eq:ExCohMag} to include the dependence of magnetic shifts on the exciton fraction $\boldsymbol{X}$,

\begin{align}
E_{pol} (t,\boldsymbol{X}) &= \boldsymbol{X} \Delta E(t) \\
&= \boldsymbol{X}\,\Delta_B \cos^2 \left( \frac{\theta}{2} \right) - \boldsymbol{X}\,\frac{\Delta_B}{2} \, \sin(\theta_0) \, \Delta \theta_0 \, \exp(-\kappa t) \, \cos(\omega t) \label{Eq:PolCohMag}\, .
\end{align}

For the simulation of the oscillatory part of the transient differential reflectance $R(t)-R_{osc}(t)$ of polaritons, we reduce \cref{Eq:PolCohMag} to $\Delta E_{pol} (t,\boldsymbol{X}) = \boldsymbol{X} \, \Delta_{\theta_{0}} \, \exp(-\kappa t) \, \cos(\omega t)$ and choose $\Delta_{\theta_{0}}=2$\,meV, $\kappa=0.5$\,ns and $\omega=2\pi f_{mag}$ with $f_{mag}=12.5$\,GHz (at $\boldsymbol{B_{ext}}=1.2$\,T, cf. \cref{fig:SFig-Experiment_pump-probe_B-field_dependence}) to match our experiments. 
To calculate the magnon-induced differences in the reflectance spectrum of excitons and polaritons shown in Fig.~3c-e of the main manuscript, we incorporate a 2\,meV-shift of the exciton resonance in the dielectric function of CrSBr and calculate the reflectance response for the different sample configurations. 
Even polaritons with moderate exciton fraction generate large magnon-induced changes in the reflectance spectrum compared to pure excitons due to an interplay of the magnon-induced energy shift, spectrally varying absorption, and small line-width of polariton modes. 

\subsection{Measurement of the coupling between polaritons and incoherent magnon}

\begin{figure*}[h!]
	\includegraphics[width=0.8\linewidth]{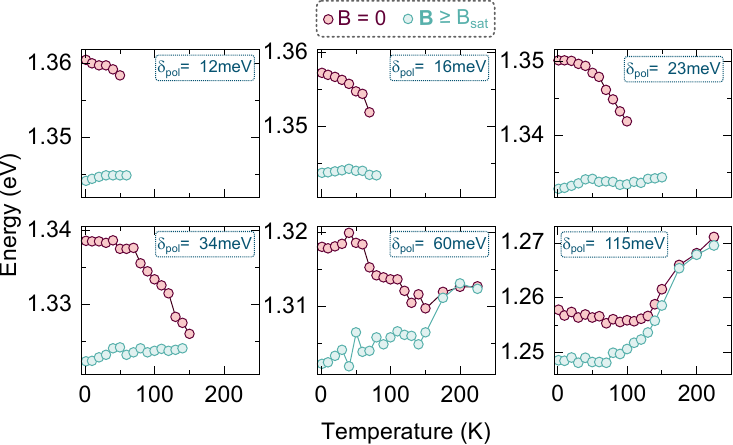}
	\caption{\textbf{Energies of different polariton branches in the 580\,nm-thin cavity sample obtained by temperature- and field-resolved PL measurements.}
		PL peak energies extracted from polariton branches with different detuning $\delta_{pol}$ for $B$=0 and $\boldsymbol{B}\ge B_{sat}$. PL signals were analyzed along the $b$--axis.
		\label{fig:SFig-Experiment_T-dependence_all-polariton-branches}}
\end{figure*}

The response of exciton-polaritons to incoherent magnons predicted by our analytic theory in \cref{Section: Theoretical analysis of the impact of coherent and incoherent magnons} is demonstrated by the temperature- and field-dependent PL emission of all polariton branches in our 580\,nm-thick cavity sample. 
Based on the results plotted in \cref{fig:SFig-Experiment_T-dependence_all-polariton-branches} we make three main observations: 
\textit{i\,)}~As a function of temperature, most exciton-polariton branches shift towards lower energies for $B=0$ and towards higher energies for $\boldsymbol{B}\geq B_{sat}$. 
\textit{ii\,)}~Polariton branches with small detuning rapidly disappear from the spectrum, while branches with large detuning are observed at much higher temperatures. 
\textit{iii\,)}~The temperature- and field-dependence of polaritons is significantly altered by detuning $\delta_{pol}$ and the exciton fraction $\boldsymbol{X}$.

\subsection{Numerical analysis of polariton-magnon coupling: Incoherent magnons}

To better understand the physical mechanisms behind these observations, we simulate the polariton temperature dependence in transfer-matrix calculations, thereby considering the influence of different parameters on the dielectric function, $\epsilon(E,T,\boldsymbol{B_{ext}})$, which is the foundation for our numerical analysis:

\begin{equation}\label{Eq:epsilon}
	\epsilon(E,T,\boldsymbol{B_{ext}})=\epsilon_{b}^\infty(T)+\frac{f_X(T)}{E_X(T,\boldsymbol{B_{ext}})^2-E^2-i\gamma(T) E} \, ,
\end{equation}

where, $E$ is the energy, $T$ the sample temperature, $\boldsymbol{B_{ext}}$ is the external magnetic field, and $f_X(T)=2\Delta_X\sqrt{{E_X(T)}^2-(\Gamma(T)/2)^2}$ denotes an \textit{effective} oscillator strength that includes the exciton energy $E_X$, the \textit{intrinsic} exciton oscillator strength $\Delta_X$, and the broadening of the exciton line-width $\Gamma(T)$. 
%The variation of $f_X(T)$ with temperature is small and thus negligible in our analysis. 

\begin{figure*}[]
	\includegraphics[width=1.0\linewidth]{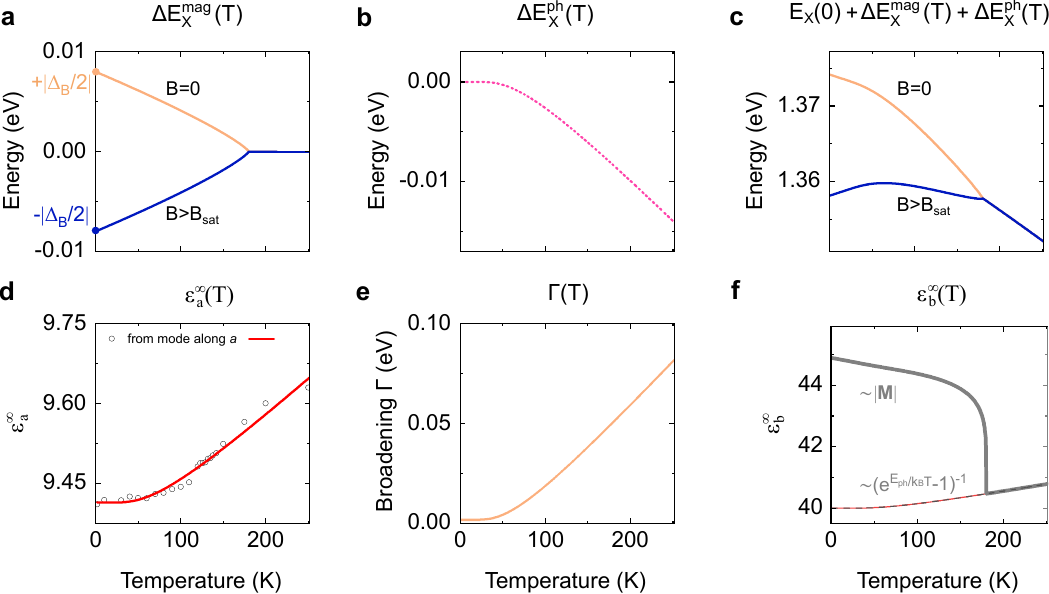}
	\caption{\textbf{Contributions to the temperature dependence of polaritons.}
		\textbf{a}~Coupling of excitons to incoherent magnons for both AFM (${B_{ext}}=0$) and FM ($\boldsymbol{B_{ext}}>B_{sat}$).
		\textbf{b}~Coupling of excitons to phonons. 
		\textbf{c}~Summed contributions of magnons and phonons. 
		\textbf{d}~Temperature-dependence of the dielectric constant $\epsilon_{a}$ measured via the temperature-dependence of the optical mode along the intermediate $a$--direction (cf. also Fig.~S3C) and fitted by the Einstein function (see text).  
		\textbf{e}~Temperature-induced broadening of the excitons line-width.
		\textbf{f}~Phenomenological temperature-dependence of the dielectric constant $\epsilon_b(\infty)$ in the magnetic easy axis direction derived from comparison with the experimental results shown in Figs. S9\,\&\,10.
		\label{fig:F9-2_Temperature_panel plot_all-contributions}}
\end{figure*}

First, we account for the different contributions to the temperature- and field-dependence of excitons in CrSBr,

\begin{equation}\label{Eq:ExTB}
E_X(T,\boldsymbol{B_{ext}})=E_X(0) + \Delta E_X^{mag} (\boldsymbol{B_{ext}},T) + \Delta E_X^{ph} (T) \,
\end{equation}

including changes in the exciton energy due to incoherent magnons, $\Delta E_X^{mag} (\boldsymbol{B_{ext}},T)$, and phonons, $\Delta E_X^{ph} (T)$.
Our numerical description of the effects of incoherent magnons is based on the results of our analytic model presented in \cref{Eq:IncMagAFM,Eq:IncMagFM}. 
We explicitly use $T_{corr}=$\,180\,K, instead of $T_N$, to account for the fact that magnetic correlations are clearly visible in our experiments up to temperatures exceeding $T_N$.
Good agreement between our numerical model and the experimental data is obtained for $\beta\approx0.4$. 
The corresponding decrease and increase of the exciton energy for AFM and FM configurations, induced by the steadily increasing density of incoherent magnons, are plotted in \cref{fig:F9-2_Temperature_panel plot_all-contributions}a. 

Besides incoherent magnons, phonons couple to excitons, inducing further temperature-dependent changes in the exciton energy that can be described by~\cite{Odonnel1991}:

\begin{equation}\label{eq:exc-phonon}
	\Delta E_X^{ph} (T) = S \langle\hbar\omega\rangle (\coth(\langle\hbar\omega\rangle)/2k_BT-1) \, ,
\end{equation}

where $S=0.5$, $\langle\hbar\omega\rangle$ is the average phonon energy (extracted from \cref{fig:F9-2_Temperature_panel plot_all-contributions}d, see discussion below), and $\hbar$ is Planck's constant. 
The sum of the terms in \cref{Eq:ExTB} is plotted in \cref{fig:F9-2_Temperature_panel plot_all-contributions}c for AFM ($B_{ext}=0$) and FM ($\boldsymbol{B_{ext}}\geq B_{sat}$) magnetic configurations. 

While the dielectric tensor element $\epsilon_a^{\infty}$ does not explicitly affect the temperature dependence of our exciton-polaritons, we can use it to experimentally access other relevant parameters.
By measuring the temperature dependence of the uncoupled optical mode in the $a$--direction in our 580\,nm cavity sample, we obtain $\epsilon_a^{\infty}(T)$, which we fit to the Einstein relation~\cite{Dubrovin2020} to extract the average optical phonon energy, $\langle\hbar\omega\rangle=$17.5\,meV, used in \cref{eq:exc-phonon}, directly from our measurements. 
This value is in reasonable agreement with the $A_g^1$ phonon mode observe in Raman experiments~\cite{Klein2022-1}.
Experimentally, we can further conclude that $\epsilon_a^{\infty}(T)$ is neither significantly affected by an external magnetic field, nor by magnetic phase transitions in CrSBr, indicating that the coupling of the magnetic order to the dielectric tensor is negligible in the $a$--direction. 

We use the extracted phonon energy to calculate the temperature-dependent broadening of the exciton line-width $\Gamma(T)$~\cite{Jin2020} (see \cref{fig:F9-2_Temperature_panel plot_all-contributions}e), which is in very good agreement with line-widths observed in PL measurements of excitons in bi- and few-layer samples~\cite{Wilson2021,Klein2022-2}. 

\begin{figure*}[h!]
	\includegraphics[width=0.8\linewidth]{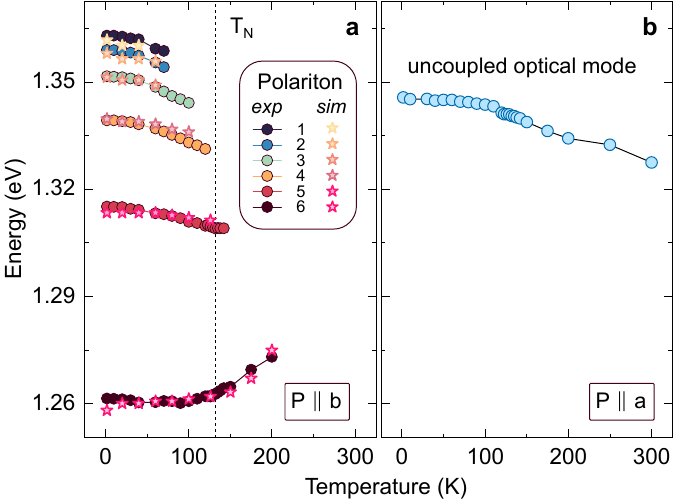}
	\caption{\textbf{Optical resonances in the 580\,nm-thin cavity sample obtained by temperature-dependent reflectivity measurements.}
		\textbf{a}~Polariton branches obtained by analyzing the reflectivity signal along the crystal $b$-direction.
		\textbf{b}~Uncoupled optical mode analyzed along the crystal $a$-direction. 
		\label{fig:SFig-Experiment_T-dependence b vs a}}
\end{figure*}

To capture the role of the polariton photon component in our simulations, we account for changes in the refractive index that determine the energy of photons in our cavities by constructing the temperature-dependence of the dielectric tensor element $\epsilon_b^{\infty}(T)$ based on the two main contributions shown in \cref{fig:F9-2_Temperature_panel plot_all-contributions}f. 
In the first step, we include the effects of phonons on the dielectric tensor element using the Einstein relation (dashed line) and the average phonon energy $\langle\hbar\omega\rangle=$17.5\,meV extracted from our measurements.
In the second step, we assume that the temperature-induced decrease of the magnetization $\boldsymbol{M}(T)$ along the magnetic easy axis ($b$--axis) reduces $\epsilon_b^{\infty}$. 
Like the field-induced splitting, we postulate that this effect occurs up to temperatures $T\approx T_{corr}$. 

In summary, considering the different contributions to $\epsilon(E,T,\boldsymbol{B_{ext}})$ shown in \cref{fig:F9-2_Temperature_panel plot_all-contributions} allows us to numerically simulate the temperature-dependence of \textit{all} polaritons branches in our 580\,nm cavity sample. 
\Cref{fig:SFig-Experiment_T-dependence b vs a} demonstrates the overall good agreement between the simulated energies and those determined by optical reflectance experiments. 
In particular, our simulations reproduce the three main conclusions deducted from our experiments: 
\textit{i\,)}~Decrease (increase) of exciton-polariton energies under increasing temperatures (shown here only for $B=0$).
\textit{ii\,)}~Rapid disappearance of the polariton branches with small detuning and high exciton fraction. 
\textit{iii\,)}~The dependence of the thermal response of exciton-polaritons on detuning and exciton fraction $\boldsymbol{X}$. 
\clearpage

\bibliography{SuppInfo.bib}
\bibliographystyle{naturemag}